\documentclass[tran]{IEEEtran}
\usepackage{array}
\usepackage{supertabular}
\usepackage{indentfirst}
\usepackage{graphicx}
\usepackage{epsfig}
\usepackage{epstopdf}
\usepackage{amsmath}
\usepackage{amsfonts}
\usepackage{subfigure}
\usepackage{color,soul}
\usepackage{cite}
\usepackage{multirow}
\begin{document}
\sethlcolor{yellow}
\title{Capacitor Voltage Synchronizing Control of 100\% Full-Scale Wind Power Generator-Supplied Power Systems}
\author{Yang Liu, \textit{Member IEEE, Member CSEE}
\thanks{Yang Liu is with the School of Electric Power Engineering, South China University of Technology, Guangzhou, China. (Email: epyangliu@scut.edu.cn)}

}

\markboth{This manuscript was written in 2016}%
{LIU \MakeLowercase{\textit{et al.}}: }

\maketitle

\begin{abstract}
This paper proposes a capacitor voltage synchronizing control (CVSC) system for the regulation of full-scale wind power generator-supplied power systems (FWPS). The capacitor combined with the inverter of a full-scale wind power generator (WPG) is controlled with a CVSC system to mimic the rotor dynamics of a synchronous generator (SG). WPGs are enabled to offer inertial response and primary regulation. The generation and load unbalance of a FWPS is reflected by the deviation of capacitor voltages of WPGs. Small-signal analysis was carried out to investigate the oscillatory modes of a FWPS with the CVSC system. Time-domain simulation studies were undertaken on the FWPS, and the performance of the CVSC system was studied in the cases where a load increase and a three-phase-to-ground fault occurred on the FWPS, respectively.
\end{abstract}

\section{Introduction}
Renewable power generation has been increasing with a record-breaking speed in the last decade. By the end of 2015, there were about 433 GW of wind power and about 231 GW of solar power spinning around the globe. The annual increase of wind power and solar power installation in 2015 was 22\% and 28.1\%, respectively \cite{windpowerreport,Solarpowerreport}. Renewable power-supplied microgrids have been considered as promising prototypes for future power supply networks \cite{5565492}. Renewable power sources are connected to power grids through flexibly controlled power electronics inverters \cite{7505630}, which introduce completely different dynamics into power grids in comparison with SGs \cite{6186805}. The stability control of future renewable power-supplied power systems is the major challenge for the development of renewable power sources \cite{7497697}. Considerate amount of research was carried out and the most widely investigated methods are the droop control-based strategies \cite{195899,1331481,4118327,7494631,1353346,4292193,5306080,7419922}.

Frequency droop control scheme was first introduced in \cite{195899} for the control of parallel-connected inverters of ac power grids operating in the stand-alone mode. The frequency and magnitude of the inverter voltage vectors were determined with active power-frequency droop and reactive power-voltage droop characteristics, respectively. The parallel-connected inverters were controlled to mimic the load sharing behavior of conventional SGs for stable frequency and voltage of the external power grid. However, this frequency and voltage droop method has a slow and oscillating transient response \cite{1331481}. As such, \cite{1331481} introduced power derivative-integral terms into the conventional frequency and voltage droop loops to improve the transient performance. Concerning the small-signal stability of such inverter-interfaced power grids with frequency droop controllers, modal analysis was undertaken in \cite{4118327} and \cite{7494631}, the results of which revealed the relationship between the oscillatory modes of the inverter-interfaced power grids and their frequency and voltage droop loops.

To overcome the slow transient response of conventional frequency droop control methods, a phase angle droop control scheme was introduced in \cite{1353346} for the control of an inverter-interfaced power grid operating in stand-alone mode. Phase angle, instead of system frequency, of the inverter voltage vector was regulated with an active power-phase angle droop characteristic to realize proper load sharing between the parallel-connected inverters. The small-signal stability of the inverter-interfaced power grids with phase angle droop controllers was studied in \cite{4292193}, which verified that high angle droop gains were required for proper load sharing under weak system conditions. However, high droop gains have a negative impact on the overall stability of the system. Hence, supplementary control loops were introduced in \cite{5306080} for the stabilization of systems with high angle droop gains. A transient stability evaluation framework was presented in \cite{7419922} for inverter-interfaced distribution systems with phase angle droop control.

For frequency droop controllers \cite{195899,1331481,4118327,7494631}, they do not require any communication between inverters, which is also their most desirable feature. Regards to angle droop controllers \cite{1353346,4292193,5306080,7419922}, they require signals from global positioning system for angle referencing, while no communication between inverters. In contrast to the above two droop control-based strategies, centralized control \cite{4582449} and master-slave control \cite{6868993} schemes were proposed for the operation of inverter-interfaced power systems as well. The centralized and master-slave control schemes need communication links between inverters, thus the stability of such power grids is influenced by the reliability of the communication links.

The existing control methods \cite{195899,1331481,4118327,7494631, 1353346,4292193,5306080,7419922,4582449, 6868993} function on the basis of that the capacitor connected to the inverter is able to offer it enough and proper amount of energy. Therefore, most of the above works \cite{195899,1331481,4118327,7494631, 1353346,4292193,5306080,7419922} have assumed that inverters are connected to ideal dc voltage sources. However, for renewable power sources, this can be hardly true without the aid of energy storages. Moreover, the beneficial dynamics of capacitors in aspects of mimicking SGs was totally ignored. The capacitors of inverters and the rotors of SGs are all energy storage devices. A capacitor can provide inertial response like a rotor as well when load changes occur in a renewable power-supplied system. Inverters regulated by conventional droop controllers actually have blocked the inertial support from capacitors.

Instead of considering inverter interfaces only, this paper uses full-scale WPGs as the power sources of renewable power-supplied power systems. The dynamics of inverters, capacitors, and energy storages are considered in the controller design process. To realize self-synchronizing and self-load sharing, a CVSC system is proposed for the operation of a FWPS. Similar to the swing equations of SGs, voltage motion equations of capacitors are used for phase angle regulation of the inverter voltage vectors of WPGs. The capacitors of WPGs absorb or release energy in a coordinated manner with the load changes of the external power gird, such that WPGs are enabled to provide inertial response. Self-load sharing is achieved by a governor designed in the CVSC system. In contrast to the active power-frequency droop characteristic, active power-capacitor voltage droop characteristic is realized with the governor.

Compared with the centralized and master-slave control strategies \cite{4582449, 6868993}, the CVSC system does not require any communication between inverters. Different from the droop control-based methods \cite{195899,1331481,4118327,7494631, 1353346,4292193,5306080,7419922}, the CVSC system enables the inverter-interfaced renewable power sources to provide inertial support to the external power grid. Attributed to the CVSC system, the FWPS investigated in this paper can operate in the same manner with conventional SG-based power systems. On this basis, the off-the-shelf results on transient stability and small-signal stability of conventional power systems can be applied easily for the FWPS.

Overall, this paper is organized as follows. Section \ref{section_description_PEIPS_investigated} presents a description of the FWPS investigated in this paper, as well as the design of the CVSC system. The results of the small-signal analysis of the FWPS are shown in Section \ref{section_small_signal_PEIPS}, and the oscillatory modes of the FWPS are investigated therein. Section \ref{section_time_domain_simulation_PEIPS} illustrates the simulation results of the FWPS obtained in two different cases, where a load increase and a three-phase-to-ground fault occurred, respectively. Based on the results of small-signal analysis and time-domain simulations, conclusions are drawn in Section \ref{sec_conclusion}, and appendices follow thereafter.
\section{Description of the FWPS and CVSC System Design}\label{section_description_PEIPS_investigated}
\subsection{Description of the FWPS Investigated in This Paper}\label{subsection_PEIPS_description}
\begin{figure}[t]
\centering
\includegraphics[width=0.49\textwidth]{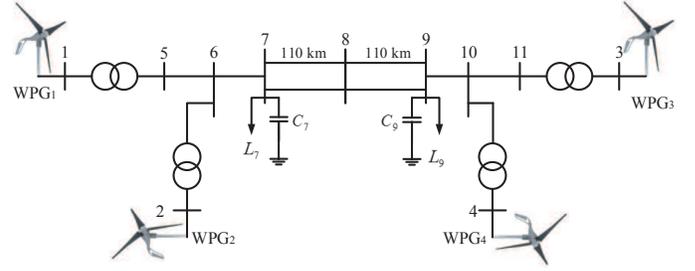}
\caption{Layout of the FWPS investigated in this paper.}
\label{fig_system_layout}
\vspace{-0.2cm}
\end{figure}

Different from the existing studies carried out on distributional-level power grids, such as \cite{1331481} and \cite{4118327}, this paper focuses on a more general application of the CVSC system, and a FWPS with high voltage level and high generation capacity is used for analysis. It should be clarified that the CVSC proposed here is feasible in distributional power grids as well. Fig. \ref{fig_system_layout} illustrates the FWPS studied in this paper, which is modified based on the Kundur four-machine two-area system \cite{kundur1994power}. The four SGs of the original system are replaced by four 889 MW wind farms, each of which is simulated with an aggregated model of a SG-based full-scale WPG having a 300 MW energy storage implemented respectively, as presented in Fig. \ref{fig_system_control}. Energy storages are connected to the capacitors of WPGs through dc-dc converters. The parameters of transformers, loads, and transmission lines are the same as those presented in \cite{kundur1994power}. Configurations of the four WPGs are the same and their parameters are presented in Appendix \ref{appen_parameter}.

\subsection{CVSC System Design}
\begin{figure}[h]
\centering
\includegraphics[width=0.49\textwidth]{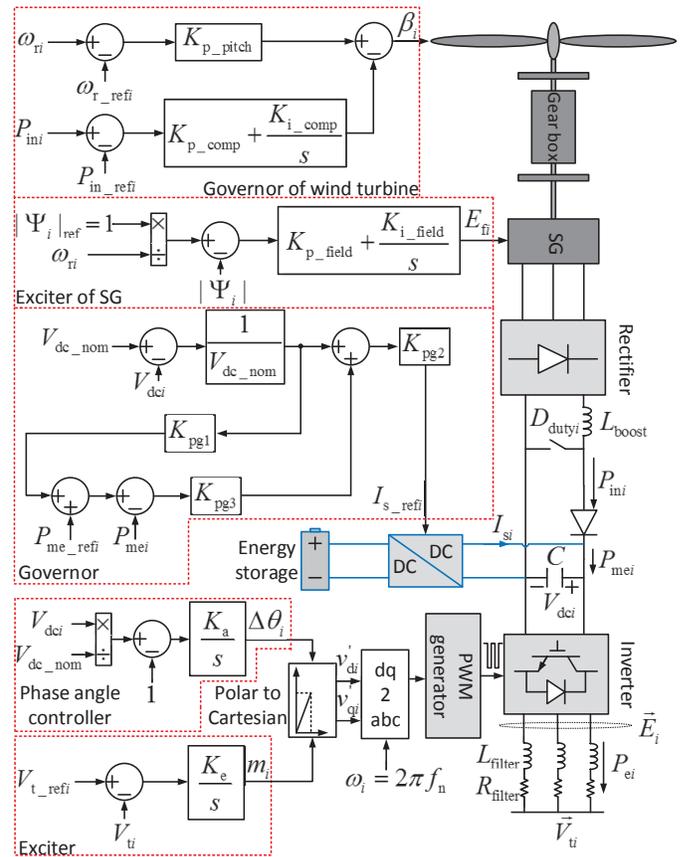}
\caption{The CVSC system of the $i$th SG-based full-scale WPG (WPG$_{i}$), where all the symbols are explained in the nomenclature presented in Appendix \ref{appen_nomenclature}.}
\label{fig_system_control}
\end{figure}
The dynamics of the capacitor voltage of WPG$_{i}$ in a FWPS can be described with
\begin{equation}\label{equ_capacitor_voltage_motion_equation}
CV_{\mathrm{dc}i}\frac{\mathrm{d}V_{\mathrm{dc}i}}{\mathrm{d}t}=P_{\mathrm{me}i}-P_{\mathrm{e}i}
\end{equation}
where $P_{\mathrm{me}i}=P_{\mathrm{in}i}+I_{\mathrm{s}i}V_{\mathrm{dc}i}$, and all the symbols shown in Fig. \ref{fig_system_control} are explained in the nomenclature of Appendix \ref{appen_nomenclature}. The rotor speed dynamics of $i$th SG in a conventional $n$-generator power system can be described with
\begin{equation}\label{equ_rotor_motion_equation}
J\omega_{\mathrm{r}i}\frac{\mathrm{d}\omega_{\mathrm{r}i}}{\mathrm{d}t}=P_{\mathrm{m}i}-P_{\mathrm{e}i}
\end{equation}
where $\omega_{\mathrm{r}i}$ represents the rotor speed of $i$th SG, $J$ denotes the inertia constant, $P_{\mathrm{m}i}$ is the mechanical power input to $i$th SG, and $P_{\mathrm{e}i}$ represents the active power output. Since the internal voltage vector of $i$th SG is inherently synchronous with its rotor speed, the rotational speed of its internal voltage vector follows (\ref{equ_rotor_motion_equation}) as well. Comparing (\ref{equ_capacitor_voltage_motion_equation}) and (\ref{equ_rotor_motion_equation}), it can be found that capacitor voltage shows the same dynamics with rotor speed when $P_{\mathrm{me}i}=P_{\mathrm{m}i}$ and $C=J$ hold. Although WPG$_{i}$ does not have a rotating mass, WPG$_{i}$ can present the same dynamics with $i$th SG in frequency-disturbed events on condition that the rotational speed of its internal voltage vector is regulated by (\ref{equ_capacitor_voltage_motion_equation}) with $P_{\mathrm{me}i}=P_{\mathrm{m}i}$ and $C=J$.

For this objective, the inverter control of WPG$_{i}$ should emulate the swing equations of $i$th SG.
The perturbed form of the swing equations of $i$th SG is
\begin{equation}\label{equ_rotor_small_signal}
\left\{
\begin{array}{l}
\displaystyle \frac{\mathrm{d}\Delta \delta_{i}}{\mathrm{d}t}=\Delta \omega_{\mathrm{r}i}\\
\displaystyle
\Delta \left(J\omega_{\mathrm{r}i}\frac{\mathrm{d}\omega_{\mathrm{r}i}}{\mathrm{d}t}\right)=\Delta P_{\mathrm{m}i}-\Delta P_{\mathrm{e}i}
\end{array}
\right.
\end{equation}
where $\Delta \delta_{i}$ is the perturbed phase angle of its internal voltage vector, and $\Delta\omega_{\mathrm{r}i}$ is the perturbed rotational speed of its internal voltage vector. Mimicking (\ref{equ_rotor_small_signal}), the perturbed phase angle $\Delta\theta_{i}$ of the inverter voltage vector of WPG$_{i}$ should be controlled according to
\begin{equation}\label{equ_capacitor_small_signal}
\left\{
\begin{array}{l}
\displaystyle \frac{\mathrm{d}\Delta \theta_{i}}{\mathrm{d}t}=K_{\mathrm{a}}\Delta V^{*}_{\mathrm{dc}i}\\
\displaystyle \Delta\left(CV_{\mathrm{dc}i}\frac{\mathrm{d}V_{\mathrm{dc}i}}{\mathrm{d}t}\right)=\Delta P_{\mathrm{me}i}-\Delta P_{\mathrm{e}i}
\end{array}
\right.
\end{equation}
where $\Delta V^{*}_{\mathrm{dc}i}=(V_{\mathrm{dc}i}-V_{\mathrm{dc}\_\mathrm{nom}})/V_{\mathrm{dc}\_\mathrm{nom}}$, and $K_{\mathrm{a}}$ is a gain designed for better transient performance. With the phase angle controller designed based on (\ref{equ_capacitor_small_signal}), the capacitor combined with the inverter of WPG$_{i}$ is enabled to offer inertial support to the FWPS as $i$th SG in the conventional SG-based power system.

To regulate the terminal voltage of WPG$_{i}$, an exciter is implemented as illustrated in Fig. \ref{fig_system_control}, where an integral loop is used to control the inverter modulation index $m_{i}$ with the voltage error measured on the point of common coupling bus (PCCB) of WPG$_{i}$. The inverter voltage vector $\overrightarrow{E}_{i}$ of WPG$_{i}$ then can be denoted as
\begin{equation}
\overrightarrow{E}_{i}=E_{i}\angle\theta_{i}=E_{i}\angle(\Delta\theta_{i}+\omega_{i}t)
\end{equation}
where $\omega_{i}=2\pi f_{\mathrm{n}}$. Then the active power output of WPG$_{i}$ in a $n$-WPG FWPS can be written as
\begin{equation}
\begin{array}{l}
P_{\mathrm{e}i}=E^2_{i}G_{ii}+\\
\sum\limits^{n}_{j=1,j\neq i}E_{i}E_{j}\left[B_{ij}\mathrm{sin}(\Delta\theta_{i}-\Delta\theta_{j})+G_{ij}\mathrm{cos}(\Delta\theta_{i}-\Delta\theta_{j})\right]
\end{array}
\end{equation}
where $G_{ij}$ and $B_{ij}$ are the real and imaginary parts of the $i$th row, $j$th column element of the admittance matrix of the FWPS. Comparatively, the active power output of $i$th SG in the conventional $n$-SG power system is \cite{kundur1994power},
\begin{equation}\label{equ_Pei_RPS}
\begin{array}{l}
P_{\mathrm{e}i}=E^2_{i}G_{ii}+\\
\sum\limits^{n}_{j=1,j\neq i}E_{i}E_{j}\left[B_{ij}\mathrm{sin}(\delta_{i}-\delta_{j})+G_{ij}\mathrm{cos}(\delta_{i}-\delta_{j})\right]
\end{array}
\end{equation}
Therefore, in the conventional SG-based power system, the stable active power output of $i$th SG implies that
\begin{equation}\label{equ_rotor_synchronize}
\frac{\mathrm{d}\delta_{i}}{\mathrm{d}t}-\frac{\mathrm{d}\delta_{j}}{\mathrm{d}t}=\omega_{i}-\omega_{j}=0
\end{equation}
where $i\neq j$. Therefore, the generation and load balance of a conventional SG-based power system is characterized by the synchronization of the rotor speed of SGs.
Referring to (\ref{equ_Pei_RPS}), the stable active power output of WPG$_{i}$ implies that
\begin{equation}\label{equ_capacitor_synchronize}
\begin{aligned}
\frac{\mathrm{d}\Delta\theta_{i}}{\mathrm{d}t}-\frac{\mathrm{d}\Delta\theta_{j}}{\mathrm{d}t}=&K_{\mathrm{a}}(\Delta V^{*}_{\mathrm{dc}i}-\Delta V^{*}_{\mathrm{dc}j})\\
=&K_{\mathrm{a}}(V^{*}_{\mathrm{dc}i}-V^{*}_{\mathrm{dc}j})=0
\end{aligned}
\end{equation}
Therefore, the generation and load balance of the FWPS is characterized by the synchronization of the capacitor voltages of WPGs on condition that $K_{\mathrm{a}}\neq 0$. Synthesizing (\ref{equ_rotor_small_signal}), (\ref{equ_capacitor_small_signal}), (\ref{equ_rotor_synchronize}), and (\ref{equ_capacitor_synchronize}), it can be found that the internal voltage vector of a WPG with the phase angle controller moves in the same manner with that of a SG. The energy stored in the capacitor of WPG$_{i}$ is
\begin{equation}\label{equ_capacitor_kinetic_energy}
E_{\mathrm{capacitor}i}=\frac{1}{2}CV^{2}_{\mathrm{dc}i}-\frac
{1}{2}CV^{2}_{\mathrm{dc0}i}=\int^{t}_{0}(P_{\mathrm{me}i}-P_{\mathrm{e}i})\mathrm{d}t
\end{equation}
where $V_{\mathrm{dc0}i}$ denotes the capacitor voltage of WPG$_{i}$ at $t=0$ s. In the case where a load increase occurs in the FWPS, capacitors release the energy stored to provide inertial support to the FWPS. The case of load drop follows similarly.

Analogous to SG-based conventional power systems, primary regulation is needed in the FWPS for proper load sharing among WPGs. A governor is implemented for capacitor voltage control as presented in Fig. \ref{fig_system_control}. The governor of a WPG controls the power flowing by the capacitor and functions after the inertial response of the capacitor. Active power-capacitor voltage droop characteristic is realized by the $K_{\mathrm{pg}1}$ loop in the governor. Compared with the case of conventional power systems, in which constant frequency errors exist after primary frequency regulations, there will be constant capacitor voltage errors after the primary regulation of WPGs in the FWPS.

With respect to the control of SGs and turbine blades of WPGs, a constant flux controller is employed for the excitation control of SGs, and a wind turbine governor is utilized for rotor speed and active power control of SGs, as shown in Fig. \ref{fig_system_control}. To verify the stability of the FWPS in the small and in the large, small-signal and time-domain simulation studies are presented in the following sections, respectively.
\section{Small-signal Analysis of the FWPS}\label{section_small_signal_PEIPS}
The FWPS introduced in Section \ref{subsection_PEIPS_description} was linearized at: $P_{\mathrm{e}1}=685$ MW, $V_{\mathrm{t}1}=1.03\angle 0^{\circ}$, $P_{\mathrm{e}2}=665$ MW, $V_{\mathrm{t}2}=1.01\angle -9.43^{\circ}$, $P_{\mathrm{e}3}=700$ MW, $V_{\mathrm{t}3}=1.03\angle -22.52^{\circ}$, $P_{\mathrm{e}4}=650$ MW, $V_{\mathrm{t}4}=1.01\angle -32.68^{\circ}$. Eighth-order models were used for the description of the SG of WPGs \cite{ong1998dynamic}, and two-mass models were employed for the drive train of wind turbines \cite{7420691}. The oscillatory modes of the FWPS are obtained as illustrated in Table \ref{Tab_model_analysis}. It can be found that all eigenvalues of the system matrix have negative real parts. Therefore, the FWPS controlled with the CVSC system is stable in the small.

Low-frequency oscillation modes, whose frequency are ranged within 0.2-2.5 Hz, are found in the rotational speed $\omega_{\mathrm{t}i} (i=1,2,3,4)$ of the wind turbine of WPGs. However, low-frequency oscillation damping is not the objective of this paper. The design of damping controllers for these low-frequency oscillatory modes has not been presented here.

\begin{figure}
\centering
\includegraphics[width=0.5\textwidth]{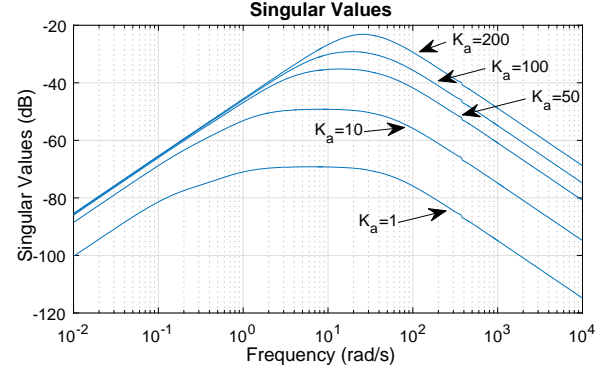}
\caption{Influence of $K_{\mathrm{a}}$ on the singular values of transfer function $\frac{P_{\mathrm{e}1}\mathrm{(p.u.)}}{V_{\mathrm{dc}1}\mathrm{(V)}}$.}
\label{fig_modes_shift}
\vspace{-0.2cm}
\end{figure}

The impact of control parameter $K_{\mathrm{a}}$ on the singular values of transfer function $\frac{P_{\mathrm{e}1}}{V_{\mathrm{dc}1}}$ is illustrated by Fig. \ref{fig_modes_shift}. It can be observed that the active power response capability of WPGs is improved as $K_{\mathrm{a}}$ increases. In turn, the influence of high frequency noise are more significant as $K_{\mathrm{a}}$ becomes larger. Considering the tradeoff between the active power response capability and the high-frequency noise attenuation, $K_{\mathrm{a}}$ is selected as 10 in this paper. Parameters of the full-scale WPGs of the FWPS presented in Appendix \ref{appen_parameter} are chosen such that all the eigenvalues of the system matrix locate in the left-half complex plane.

\begin{table}[t]
\centering
\caption{\label{Tab_model_analysis}System Modes of the RPS Controlled by CVSC Scheme}
\begin{tabular}{p{0.6cm} p{1.5cm}p{1.2cm}p{1.cm}p{1cm}p{0.9cm}}
\hline
\multicolumn{3}{c}{Eigenvalues} & Frequency & Damping & Dominant  \\
\cline{1-3}
No. & Real & Imaginary & (Hz) & Ratio & States \\
\hline
1,2 & -9.0414 & $\pm$3.7606E2 & 59.8521 & 0.0240 &  $\Psi_{\mathrm{q}3}$ \\
3,4 & -9.0234 & $\pm$3.7598E2 & 59.8388 & 0.0240 & $\Psi_{\mathrm{q}1}$ \\
5,6 & -9.0232 & $\pm$3.7597E2 & 59.8379 & 0.0240 & $\Psi_{\mathrm{q}2}$ \\
7,8 & -9.0277 & $\pm$3.7597E2 & 59.8372 & 0.0240 & $\Psi_{\mathrm{q}4}$ \\
9 & -52.9610 & - & - & - & $V_{\mathrm{dc}1}$ \\
10 & -52.7246 & - & - & - & $V_{\mathrm{dc}3}$\\
11 & -51.6736 & - & - & - & V$_{\mathrm{dc}2}$ \\
12 & -51.6517 & - & - & - & V$_{\mathrm{dc}4}$ \\
13,14 & -5.4940 & $\pm$28.5222 & 4.5395 & 0.1891 & $\omega_{\mathrm{r}1}$\\
15,16 & -5.6389 & $\pm$30.7479 & 4.8937 & 0.1804 & $\omega_{\mathrm{r}4}$\\
17,18 & -5.6614 & $\pm$28.9264 & 4.6038 & 0.1921 & $\omega_{\mathrm{r}2}$ \\
19,20 & -5.5174 & $\pm$29.8855 & 4.7564 & 0.1816 & $\omega_{\mathrm{r}3}$ \\
21 & -29.9355 & - & - & - & $\Psi^{'}_{\mathrm{kq}1}$ \\
22 & -29.3762 & - & - & - & $\Psi^{'}_{\mathrm{kq}2}$ \\
23 & -29.5797 & - & - & - & $\Psi^{'}_{\mathrm{kq}3}$ \\
24 & -29.0189 & - & - & - & $\Psi^{'}_{\mathrm{kq}4}$ \\
25 & -16.1247 & - & - & - & $\Psi^{'}_{\mathrm{kd}1}$ \\
26 & -16.6668 & - & - & - & $\Psi^{'}_{\mathrm{kd}4}$ \\
27 & -16.3040 & - & - & - & $\Psi^{'}_{\mathrm{kd}2}$ \\
28 & -16.4182 & - & - & - & $\Psi^{'}_{\mathrm{kd}3}$ \\
29,30 & -0.1282 & $\pm$3.2972 & 0.5248 & 0.0388 & $\omega_{\mathrm{t}1}$ \\
31,32 & -0.1200 & $\pm$3.3042 & 0.5259 & 0.0363 & $\omega_{\mathrm{t}2}$ \\
33,34 & -0.1032 & $\pm$3.3433 & 0.5321 & 0.0309 & $\omega_{\mathrm{t}4}$\\
35,36 & -0.1124 & $\pm$3.3273 & 0.5296 & 0.0337 & $\omega_{\mathrm{t}3}$\\
37 & -1.3278 & - & - & - & $\Delta\theta_{4}$\\
38 & -1.3049 & - & - & - & $\Delta\theta_{2}$ \\
39 & -8.9819E-13 & - & - & - & $\Delta\theta_{1}$ \\
40 & -0.6776 & - & - & - & $\Psi^{'}_{\mathrm{f}4}$ \\
41 & -0.5778 & - & - & - & $\Psi^{'}_{\mathrm{f}3}$ \\
42 & -0.5318 & - & - & - & $\Psi^{'}_{\mathrm{f}2}$ \\
43,44 & -0.2934 & $\pm$0.1063 & 0.0169 & 0.9402 & $m_{2}$ \\
45 & -0.4442 & - & - & - & $\Psi^{'}_{\mathrm{f}1}$\\
46 & -0.3810 & - & - & - & $m_{3}$ \\
47 & -0.2485 & - & - & - & $m_{1}$\\
48 & -0.2465 & - & - & - & $m_{4}$ \\
\hline
\end{tabular}
\vspace{-0.2cm}
\end{table}

\section{Time-domain Simulation Results of the FWPS}\label{section_time_domain_simulation_PEIPS}
In order to verify the stability of the FWPS in the large, time-domain simulations considering electromagnetic transients of the system were undertaken, and the obtained results are presented in this section. In the FWPS, WPG$_{2}$ and WPG$_{4}$ are implemented with the CVSC system. WPG$_{3}$ is chosen as the slake machine, and it does not have a governor installed.
\subsection{A 400 MW Load Increase Occurred on Load Bus 9 of the FWPS at $t=2$ s}
\begin{figure}[!h]
\centering
\includegraphics[width=0.48\textwidth]{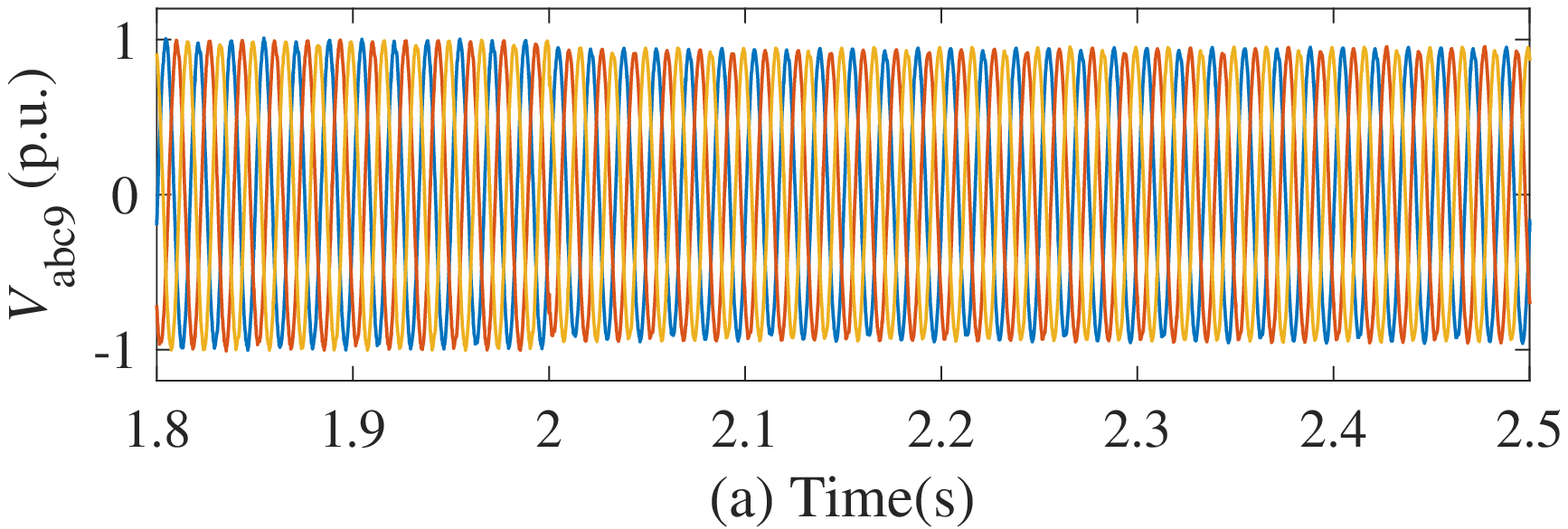}\\
\includegraphics[width=0.48\textwidth]{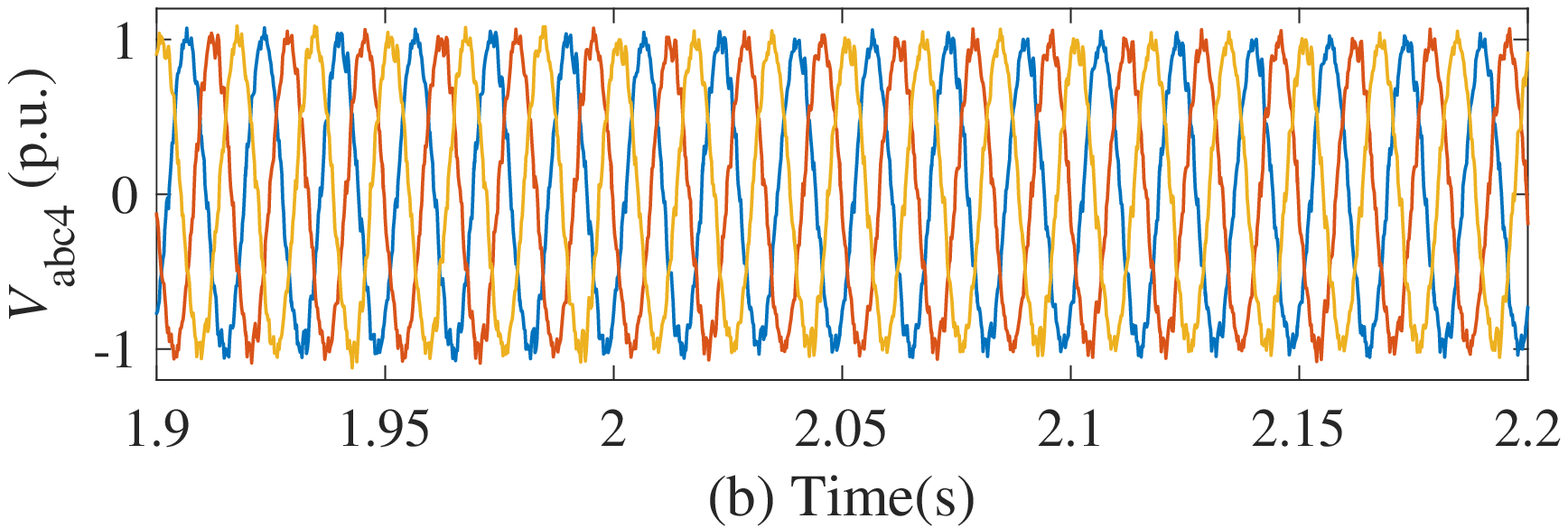}\\
\includegraphics[width=0.48\textwidth]{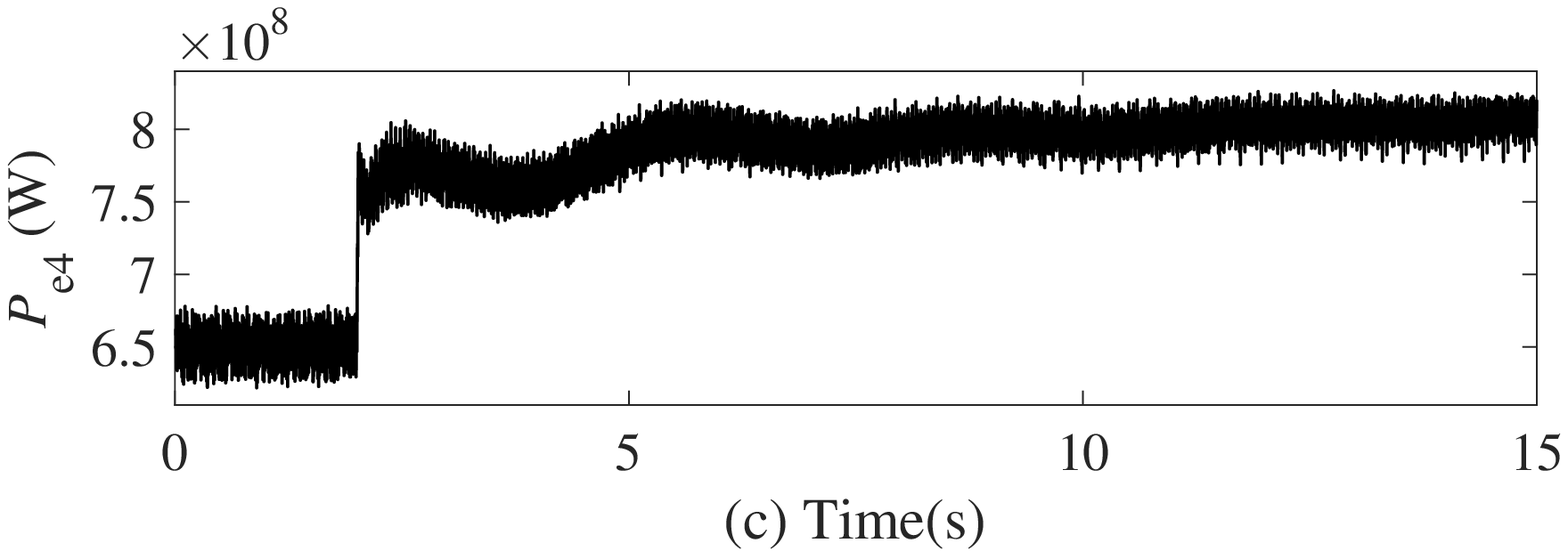}\\
\includegraphics[width=0.48\textwidth]{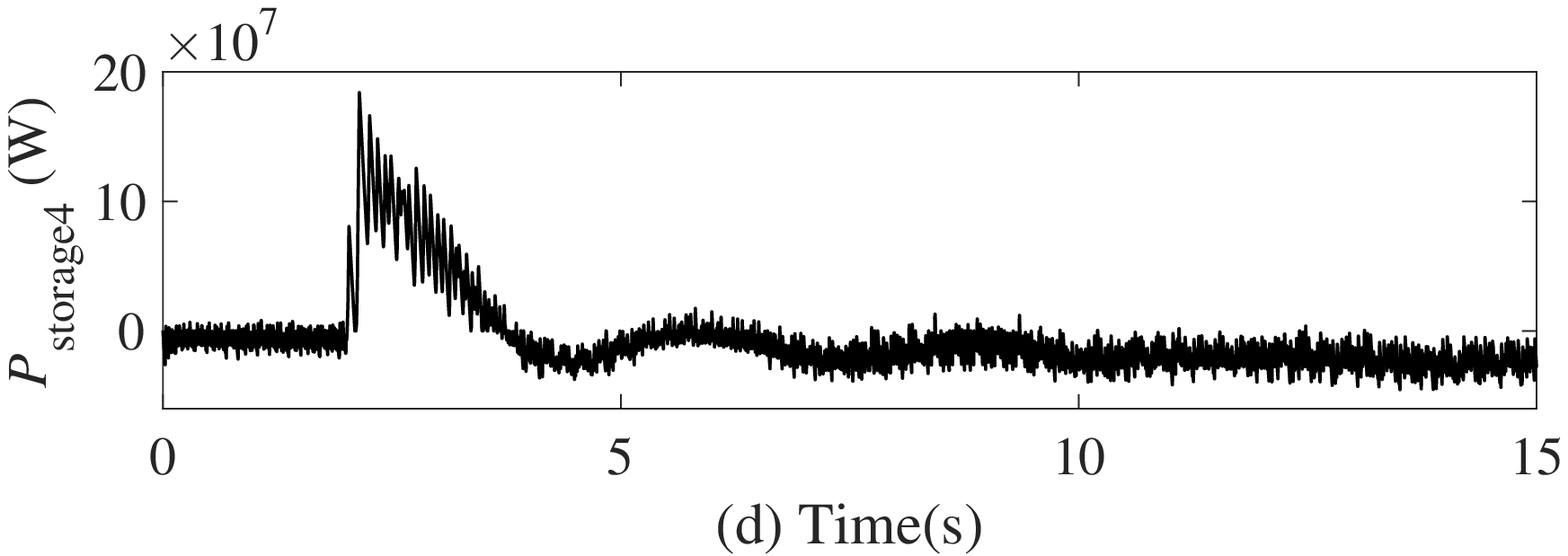}\\
\includegraphics[width=0.48\textwidth]{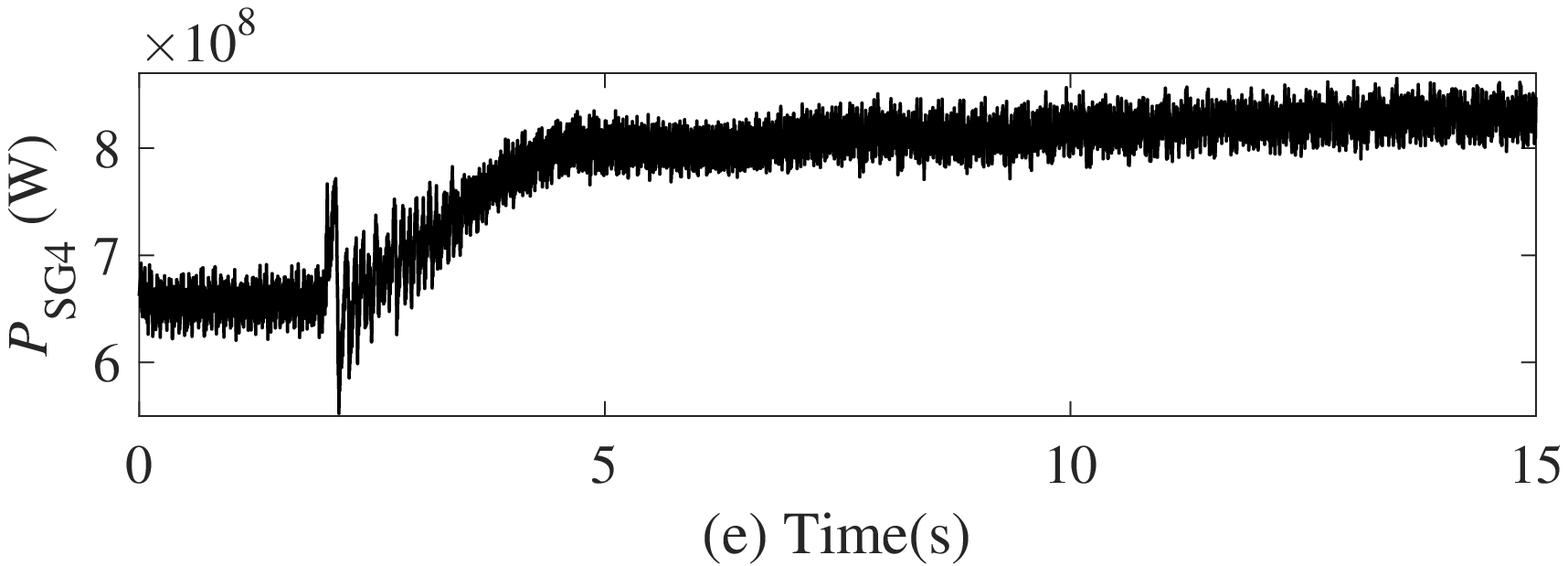}\\
\includegraphics[width=0.48\textwidth]{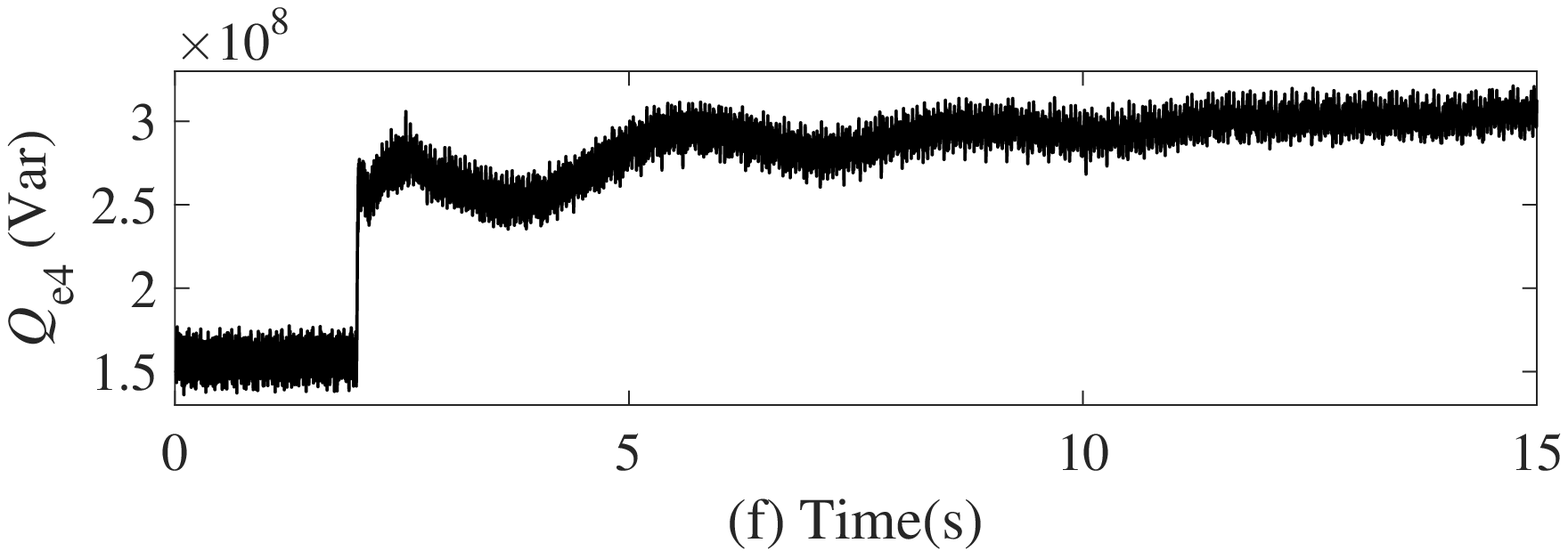}\\
\includegraphics[width=0.48\textwidth]{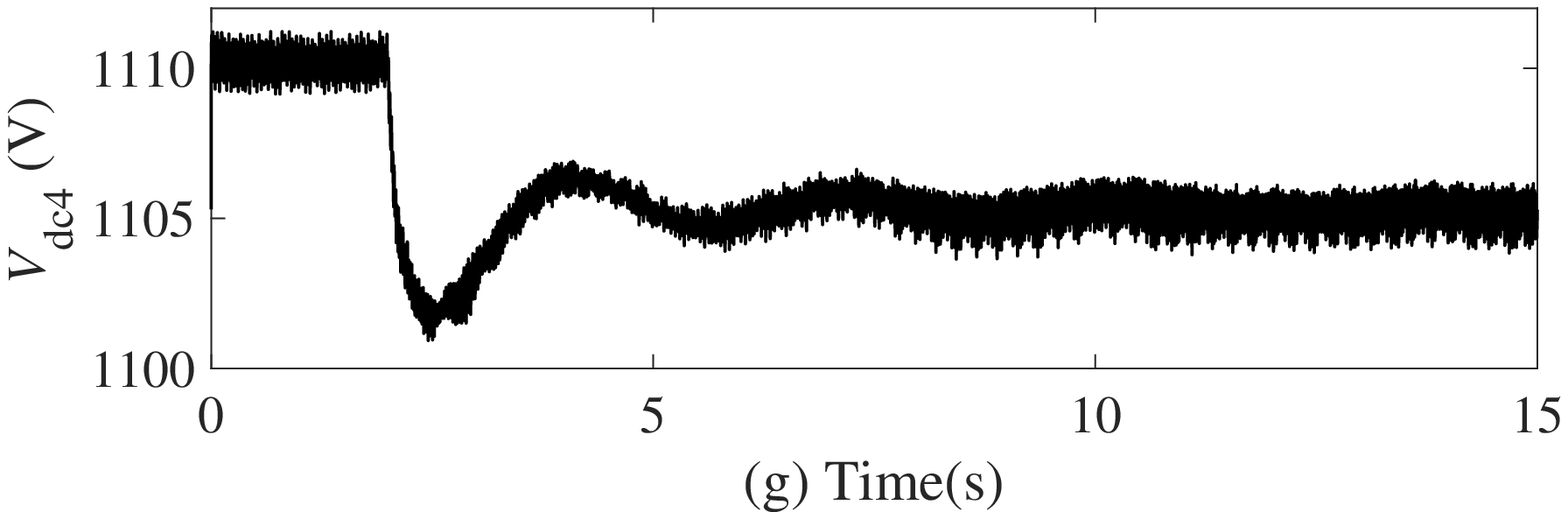}
\caption{Dynamics of WPG$_{4}$ obtained in the case where a 400 MW load was connected on load bus 9 at $t=2$ s. ((a) Three-phase voltages measured on load bus 9 (b) Three-phase voltages measured on generator bus 4 (c) Active power output of WPG$_{4}$ (d) Power output of the energy storage of WPG$_{4}$ (e) Active power output of the SG of WPG$_{4}$ (f) Reactive power output of WPG$_{4}$ (g) Capacitor voltage of WPG$_{4}$)}
\label{fig_load_change1}
\vspace{-0.2cm}
\end{figure}

A 400 MW load was connected on load bus 9 at $t=2$ s. The dynamics of WPG$_{4}$ and that of the load bus voltage are presented in Fig. \ref{fig_load_change1}. Due to the load increase, the magnitude of load bus voltage dropped, and the three-phase voltages measured on load bus 9 are shown in Fig. \ref{fig_load_change1} (a). Meanwhile, the load current increased, which resulted in more active power output of WPG$_{4}$ as illustrated in Fig. \ref{fig_load_change1} (c). This process was the inertial response from WPG$_{4}$, which was achieved by the combined effort of the capacitor and the energy storage. As depicted in Fig. \ref{fig_load_change1} (g), the capacitor voltage of WPG$_{4}$ dropped due to the upsurge of its active power output. Moreover, the energy storage provided more power to the capacitor as illustrated in Fig. \ref{fig_load_change1} (d). The drop of capacitor voltage also lead to the active power output increase of the SG of WPG$_{4}$, which is verified by Fig. \ref{fig_load_change1} (e). In this way, the kinetic energy stored in the rotor of the SG was released to provided inertial support for the capacitor voltage drop. Besides the increase of active power generation, WPG$_{4}$ generated more reactive power to the external power grid, and a step increase of reactive power can be observed as shown in Fig. \ref{fig_load_change1} (f). As a result, the PCCB voltage of WPG$_{4}$ was maintained as presented in Fig. \ref{fig_load_change1} (b).

\begin{figure}[!h]
\centering
\includegraphics[width=0.48\textwidth]{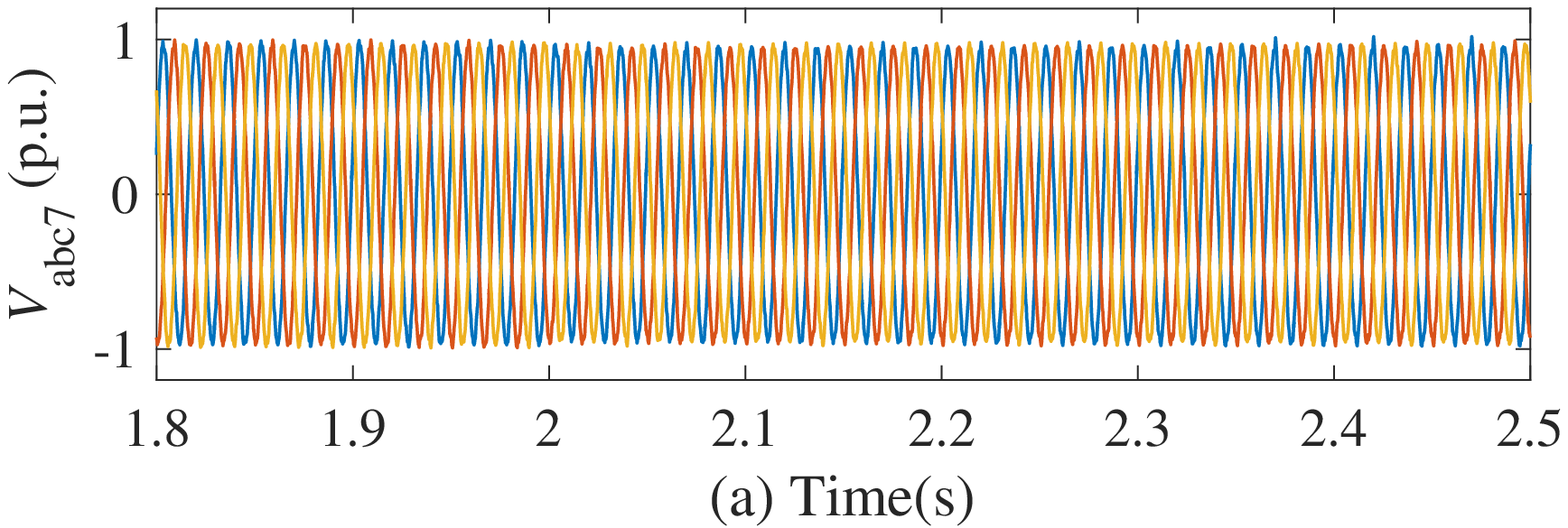}\\
\includegraphics[width=0.48\textwidth]{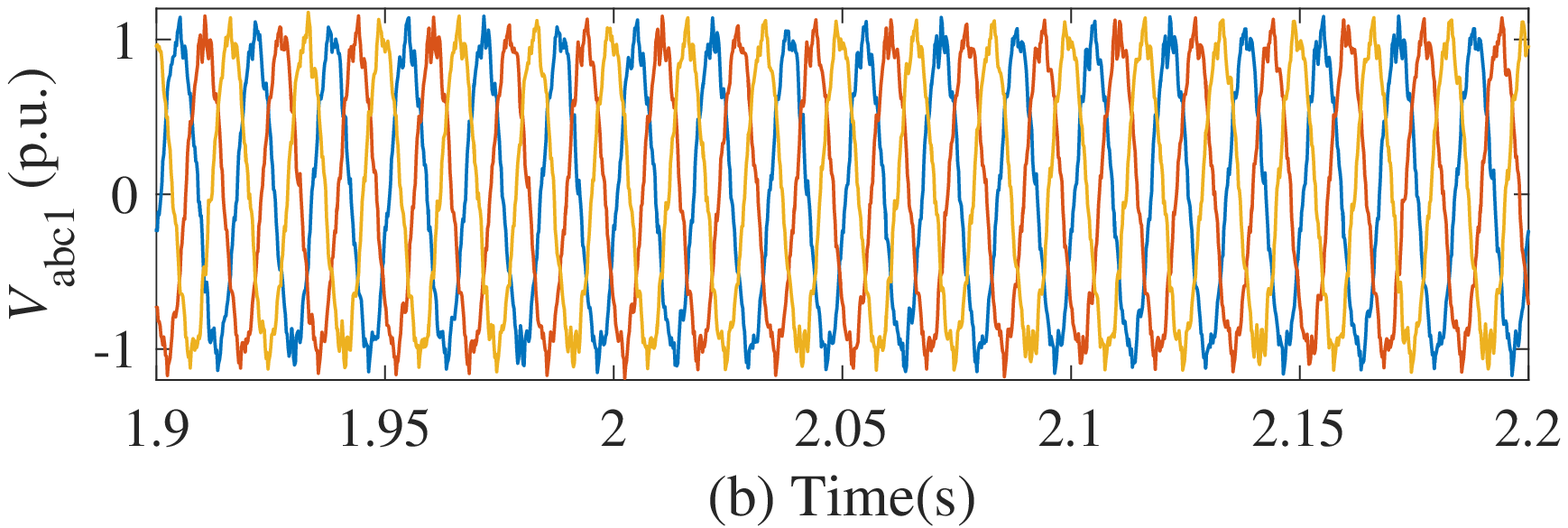}\\
\includegraphics[width=0.48\textwidth]{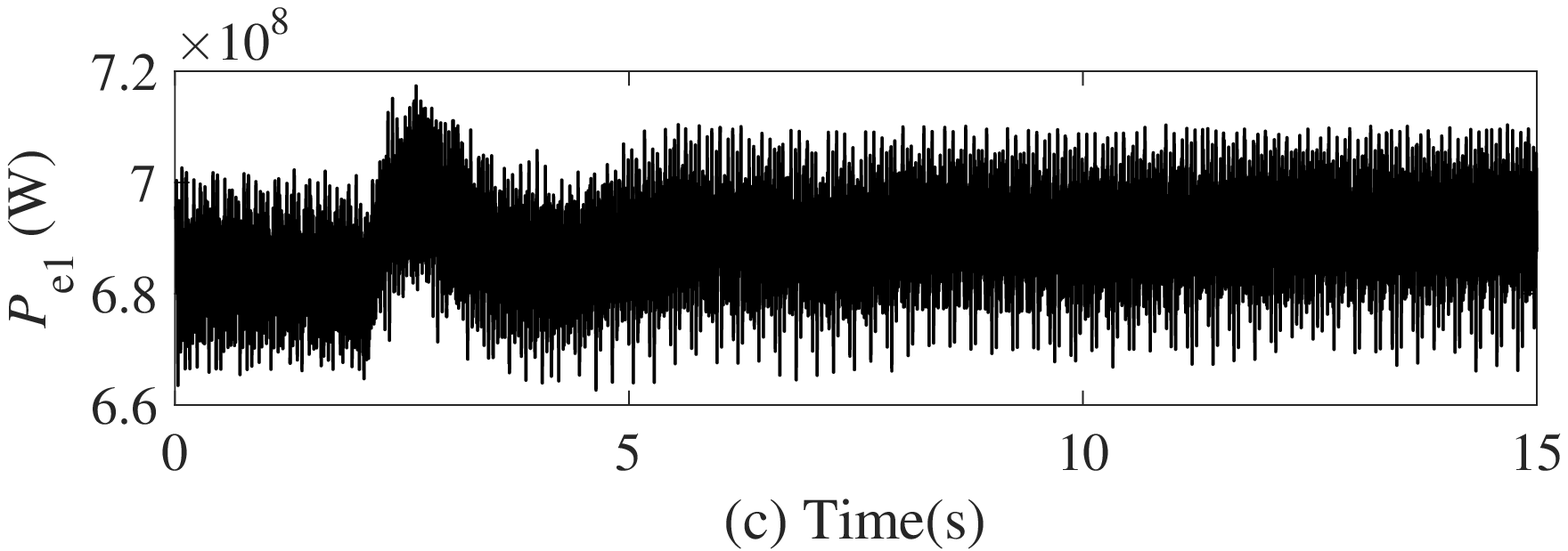}\\
\includegraphics[width=0.48\textwidth]{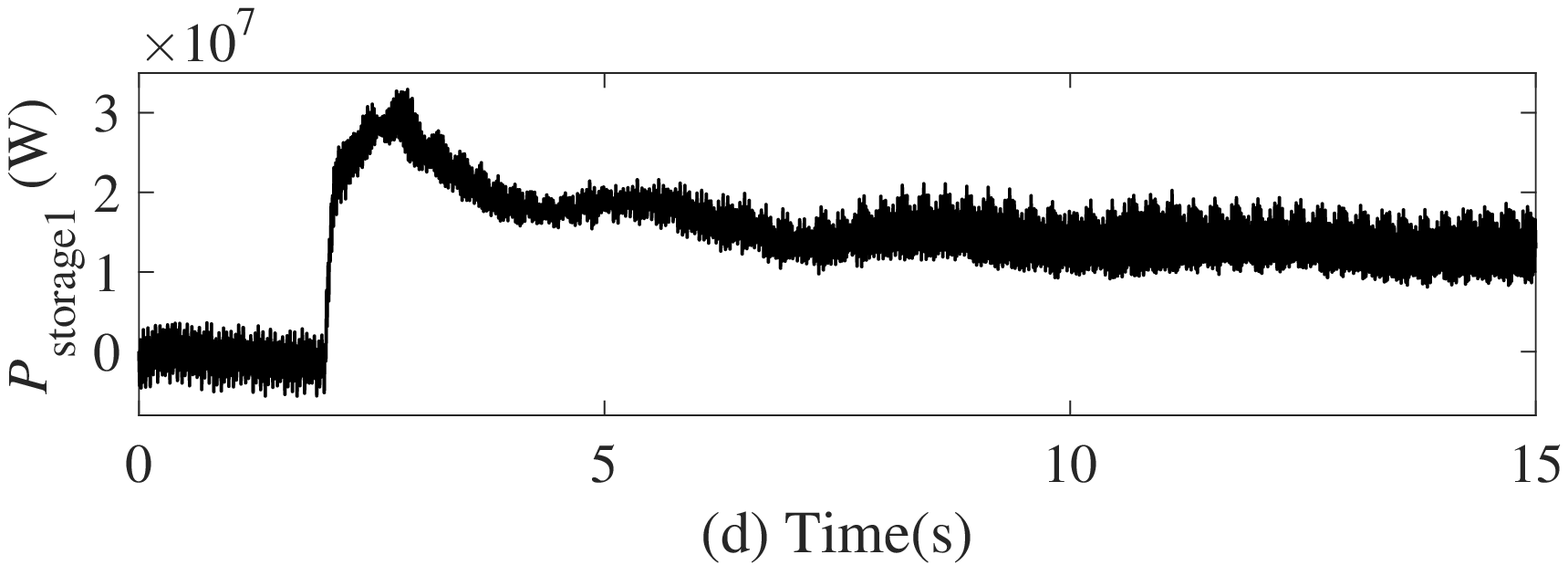}\\
\includegraphics[width=0.48\textwidth]{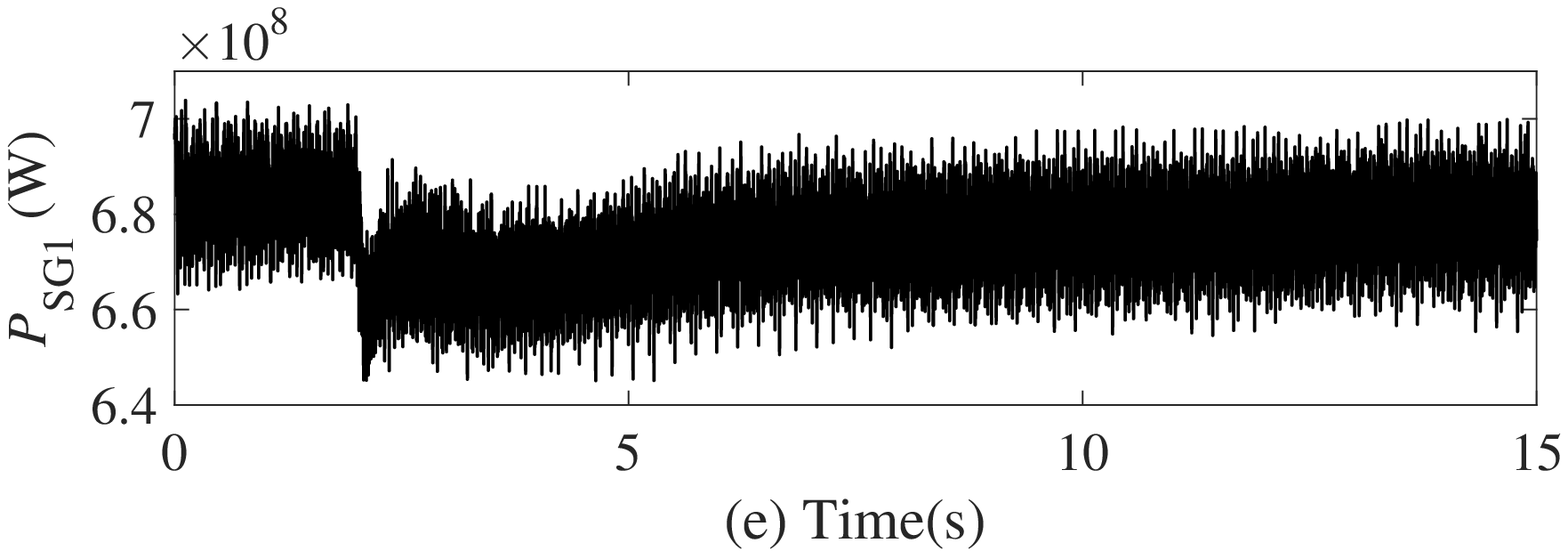}\\
\includegraphics[width=0.48\textwidth]{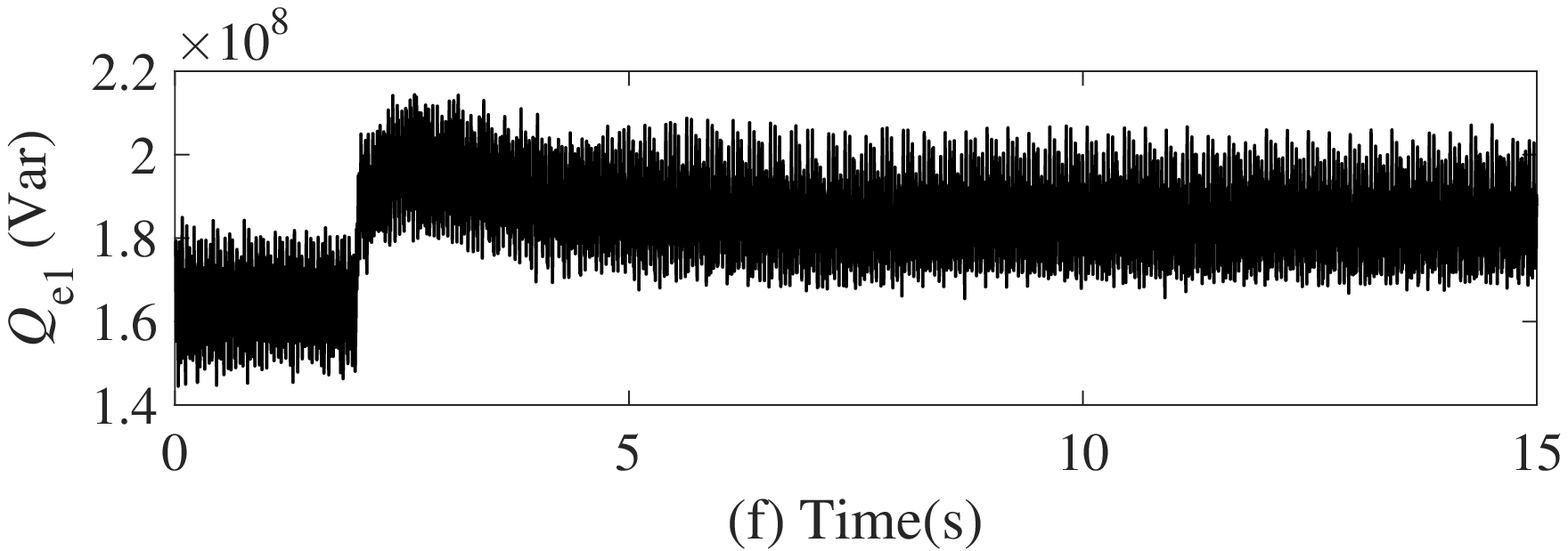}\\
\includegraphics[width=0.48\textwidth]{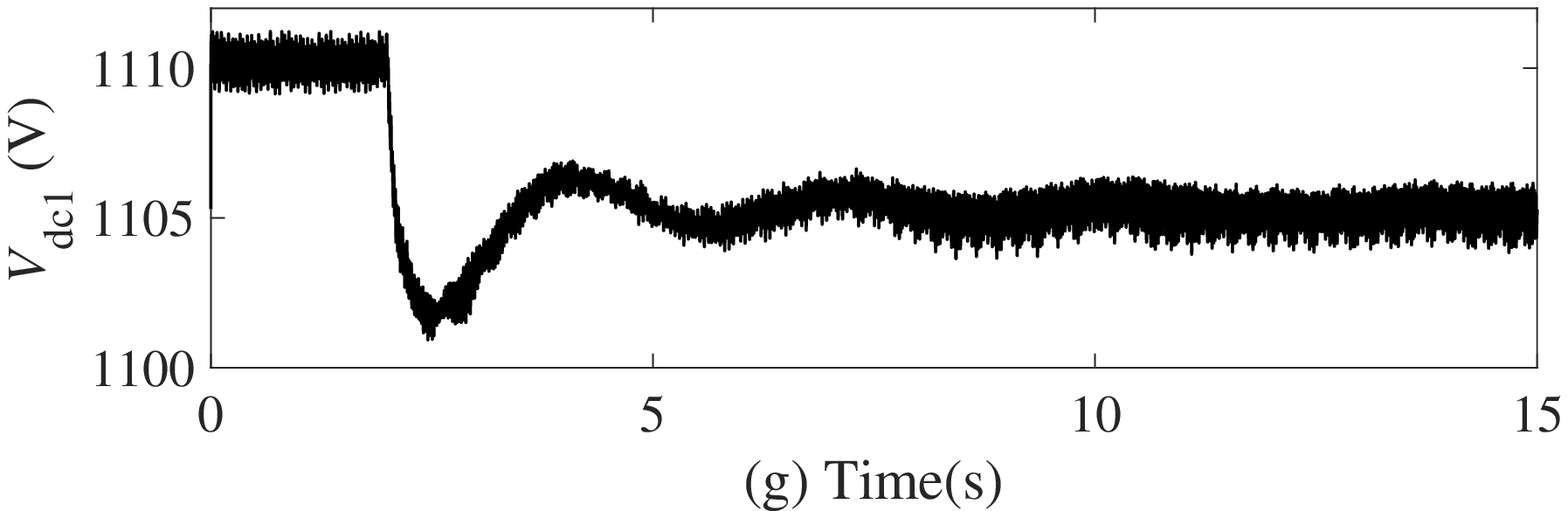}
\caption{Dynamics of WPG$_{1}$ obtained in the case where a 400 MW load was connected on load bus 9 at $t=2$ s. ((a) Three-phase voltages measured on load bus 7 (b) Three-phase voltages measured on generator bus 1 (c) Active power output of WPG$_{1}$ (d) Power output of the energy storage of WPG$_{1}$ (e) Active power output of the SG of WPG$_{1}$ (f) Reactive power output of WPG$_{1}$ (g) Capacitor voltage of WPG$_{1}$)}
\label{fig_load_change2}
\vspace{-0.2cm}
\end{figure}

\begin{figure}[ht]
\centering
\includegraphics[width=0.48\textwidth]{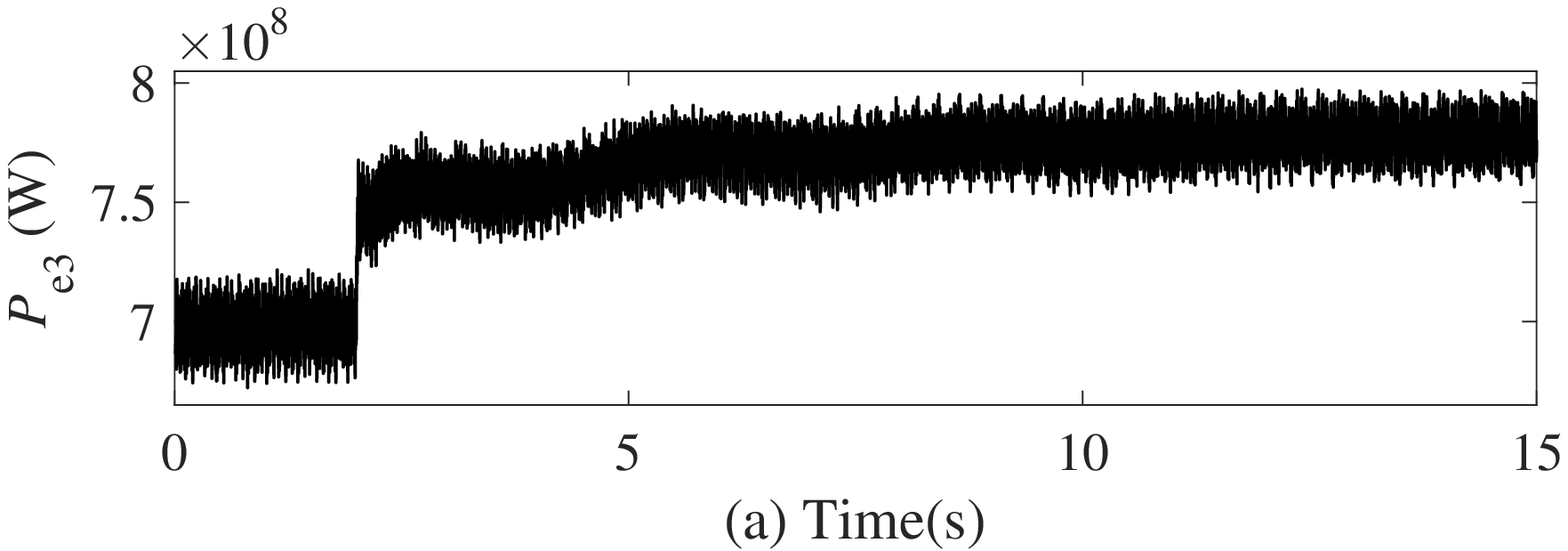}\\
\includegraphics[width=0.48\textwidth]{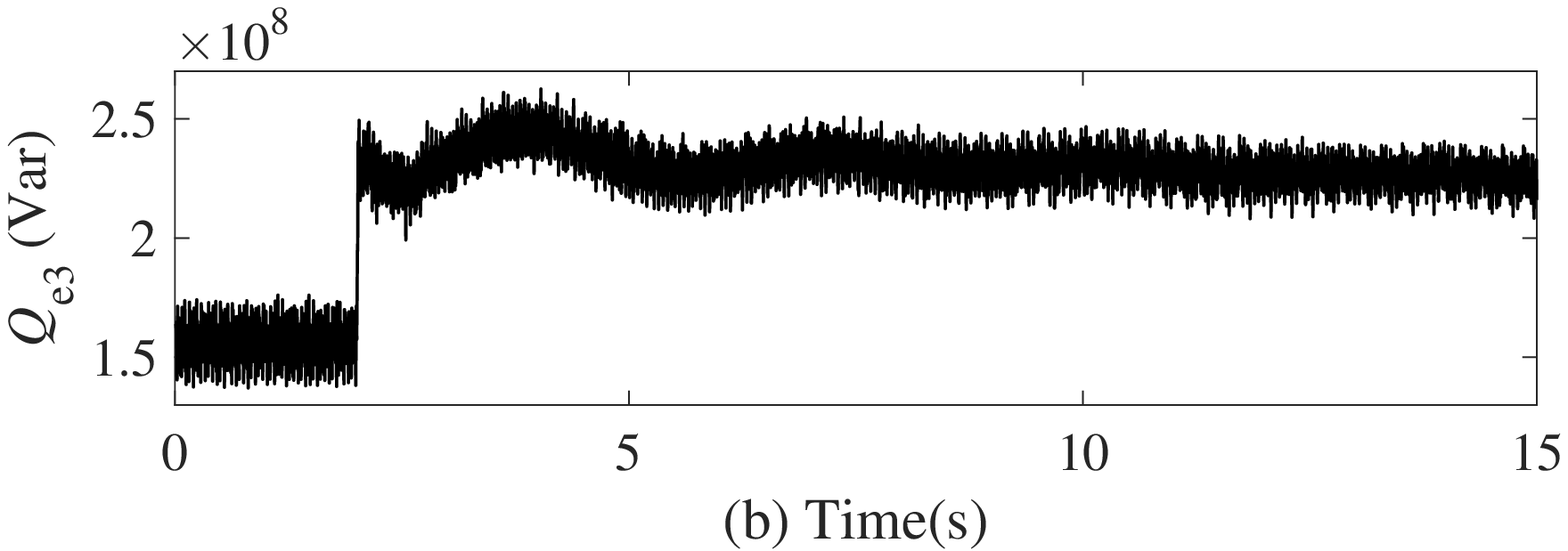}\\
\includegraphics[width=0.48\textwidth]{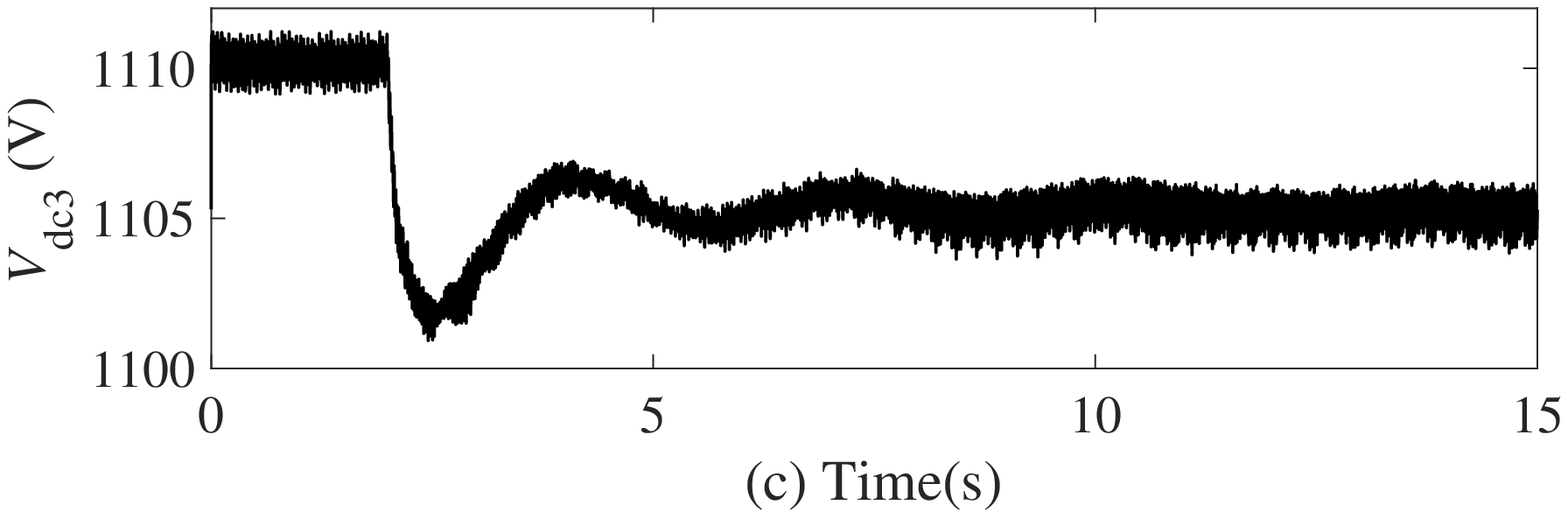}\\
\includegraphics[width=0.48\textwidth]{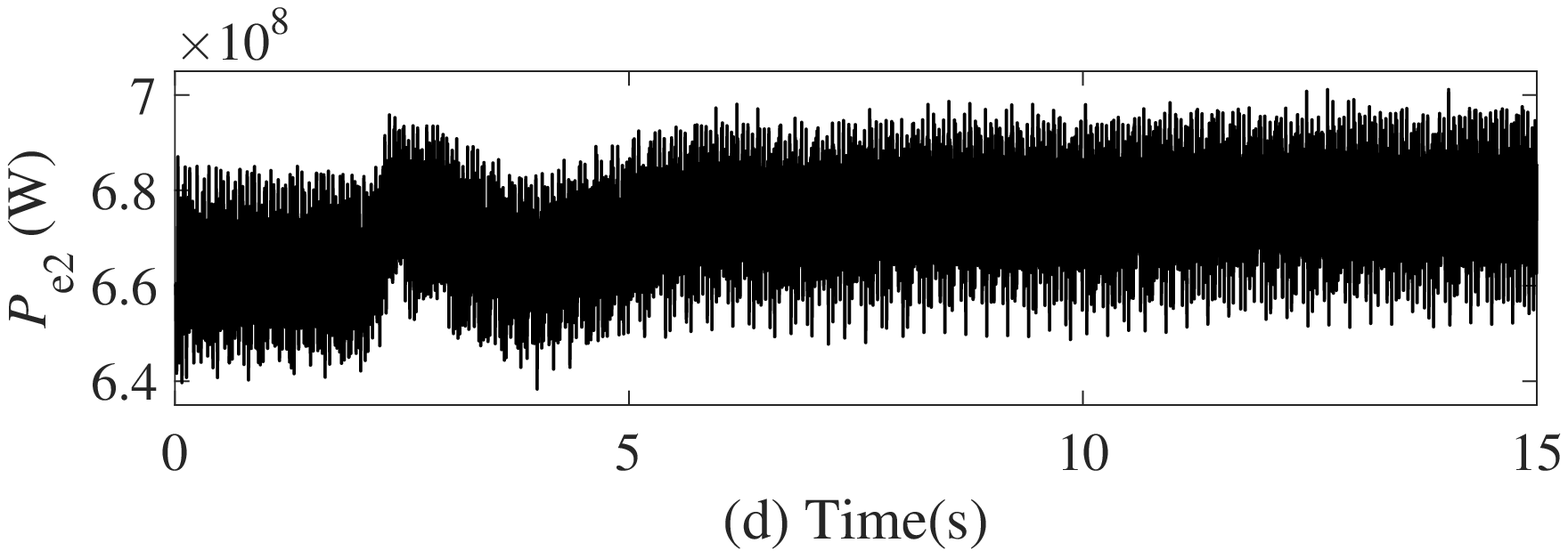}\\
\includegraphics[width=0.48\textwidth]{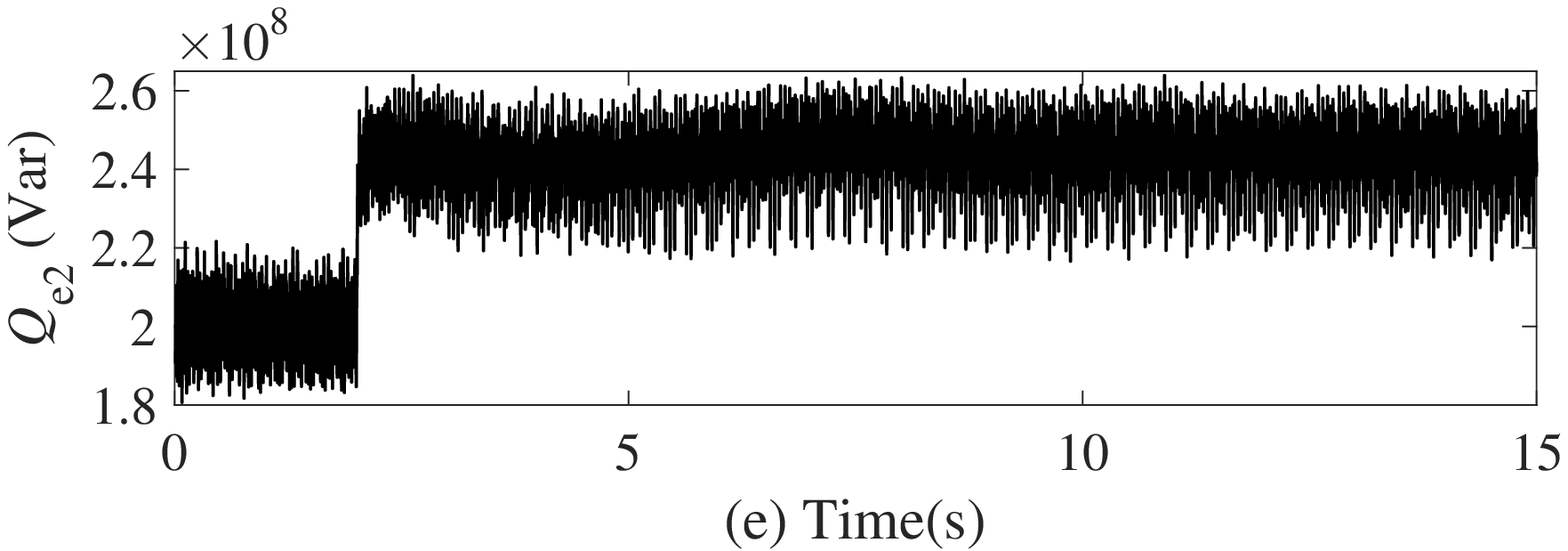}\\
\includegraphics[width=0.48\textwidth]{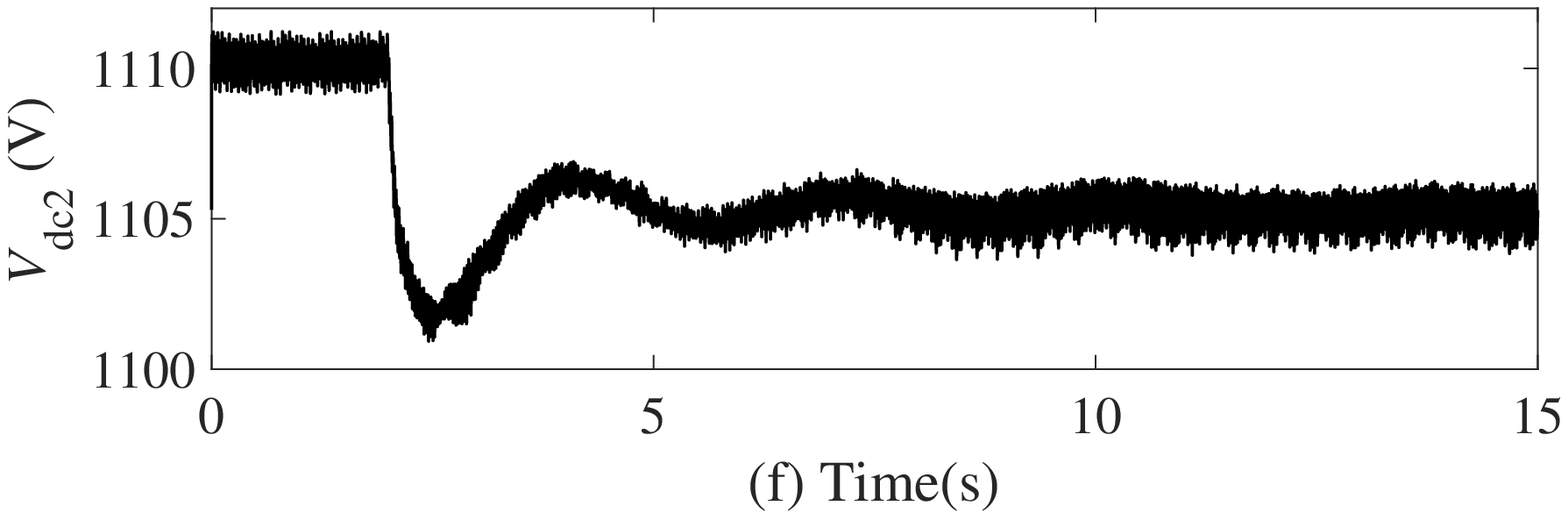}
\caption{Dynamics of WPG$_{2}$ and WPG$_{3}$ obtained in the case where a 400 MW load was connected on load bus 9 at $t=2$ s. ((a) Active power output of WPG$_{3}$ (b) Reactive power output of WPG$_{3}$ (c) Capacitor voltage of WPG$_{3}$ (d) Active power output of WPG$_{2}$ (e) Reactive power output WPG$_{2}$ (f) Capacitor voltage of WPG$_{2}$)}
\label{fig_load_change3}
\vspace{-0.6cm}
\end{figure}
In contrast to load bus 9, the voltage of load bus 7 located at the other side of the FWPS presented less magnitude drop as shown in Fig. \ref{fig_load_change2} (a). Therefore, the generation current increase of WPG$_{1}$ was smaller than that of WPG$_{4}$, which is illustrated by the active power output of WPG$_{1}$ depicted in Fig. \ref{fig_load_change2} (c). The dynamics of the capacitor voltage of WPG$_{1}$ is shown in Fig. \ref{fig_load_change2} (g). The capacitor voltage of WPG$_{1}$ synchronized with that of WPG$_{4}$, which matches the analysis presented in Section \ref{section_description_PEIPS_investigated}-B. Due to the capacitor voltage drop, the energy storage of WPG$_{1}$ increased its power output as illustrated in Fig. \ref{fig_load_change2} (d). Meanwhile, the active power output of the SG of WPG$_{1}$ decreased slightly as depicted in Fig. \ref{fig_load_change2} (e), in a coordinated manner with the energy storage dynamics such that $P_{\mathrm{e}1}$ presented in Fig. \ref{fig_load_change2} (c) was obtained. The step up of the reactive power output of WPG$_{1}$ is presented in Fig. \ref{fig_load_change2} (f). The three-phase voltages depicted in Fig. \ref{fig_load_change2} (b) were measured on the PCCB of WPG$_{1}$.

\begin{figure}[!ht]
\centering
\includegraphics[width=0.485\textwidth]{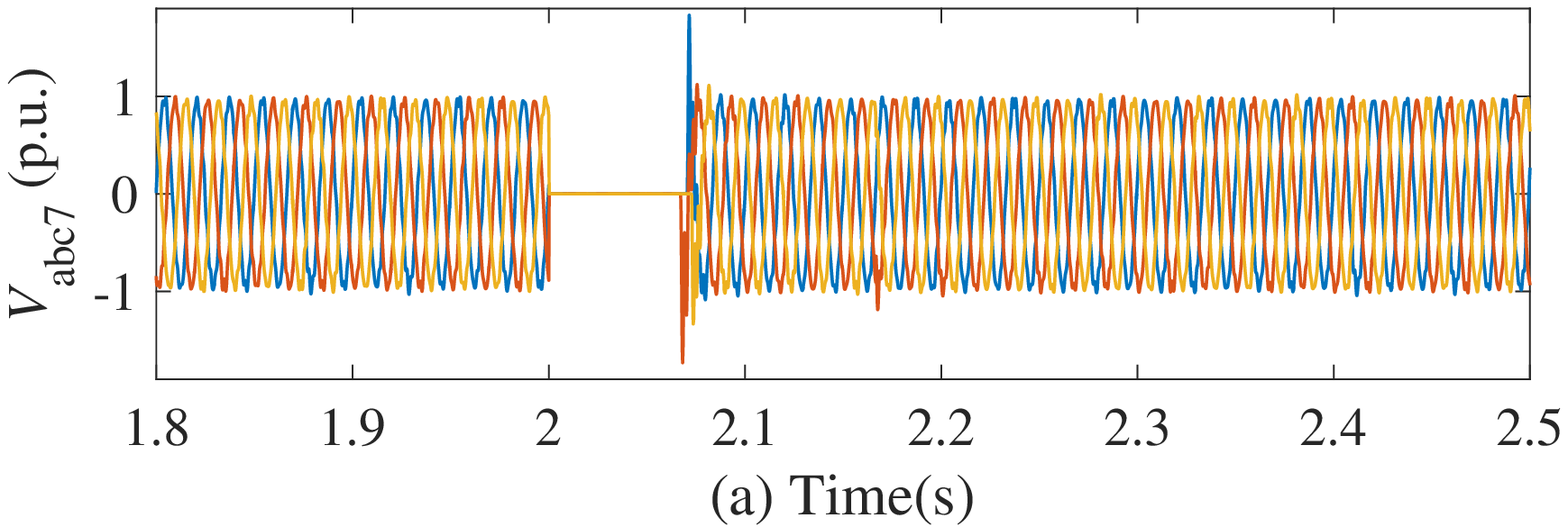}\\
\includegraphics[width=0.485\textwidth]{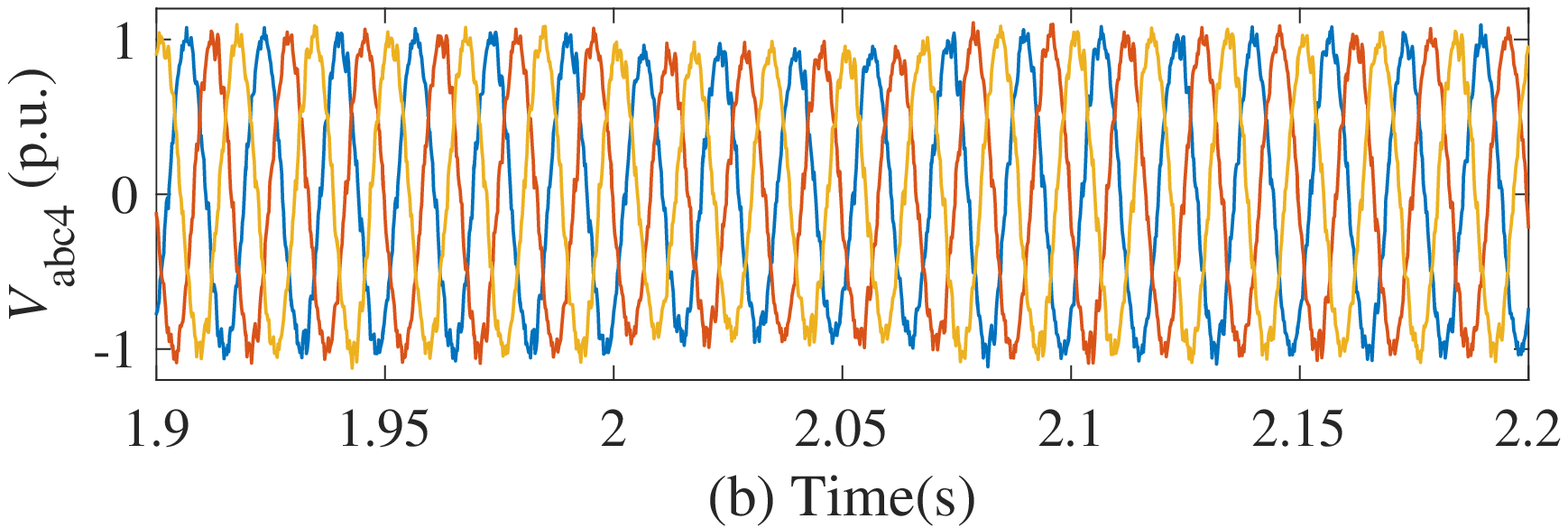}\\
\includegraphics[width=0.485\textwidth]{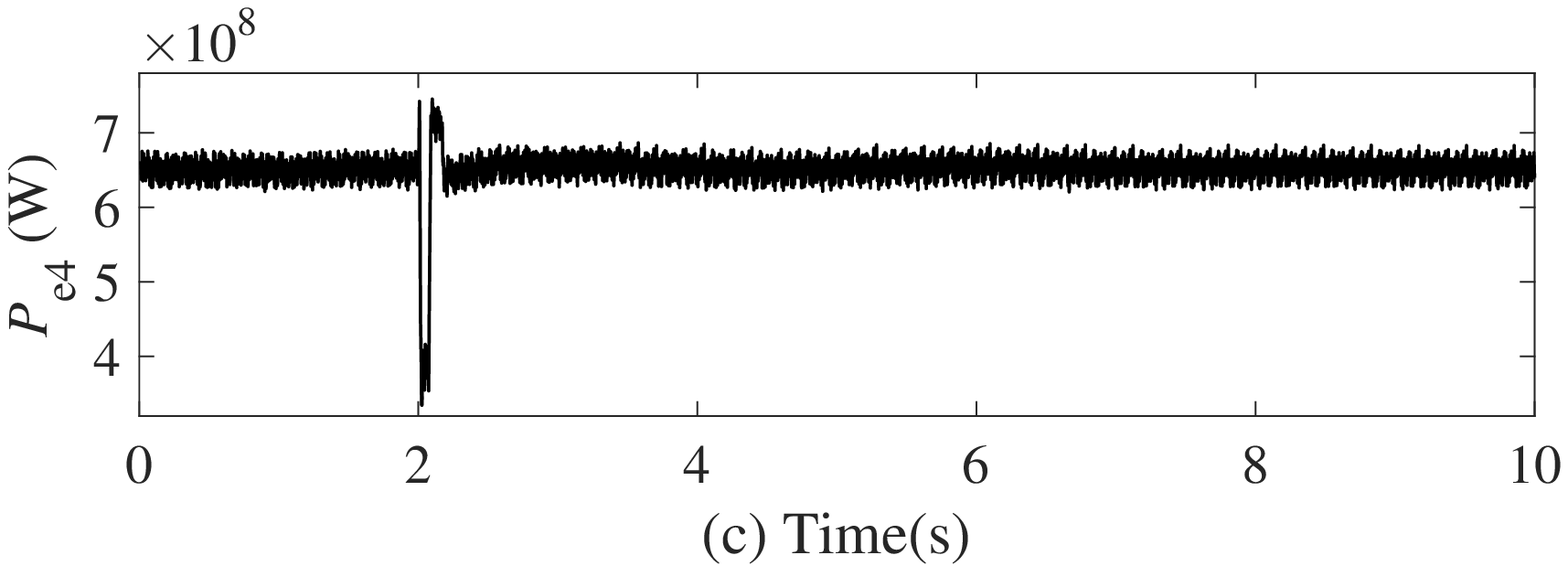}\\
\includegraphics[width=0.485\textwidth]{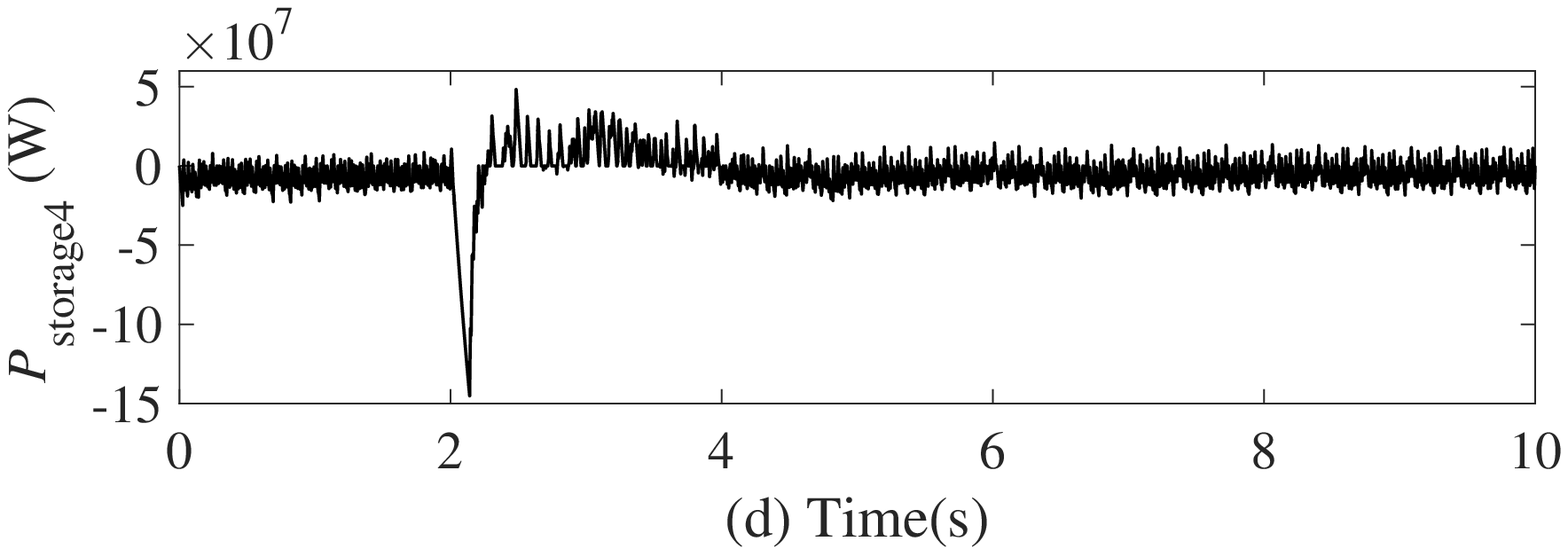}\\
\includegraphics[width=0.485\textwidth]{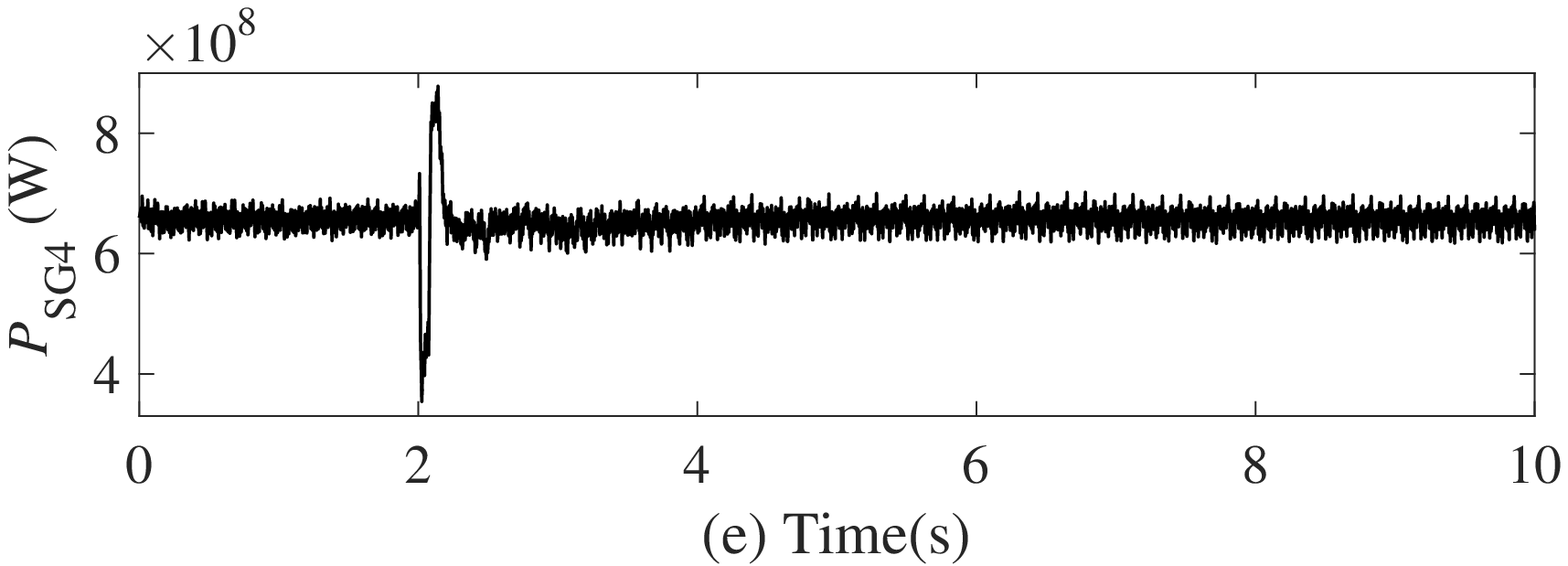}\\
\includegraphics[width=0.485\textwidth]{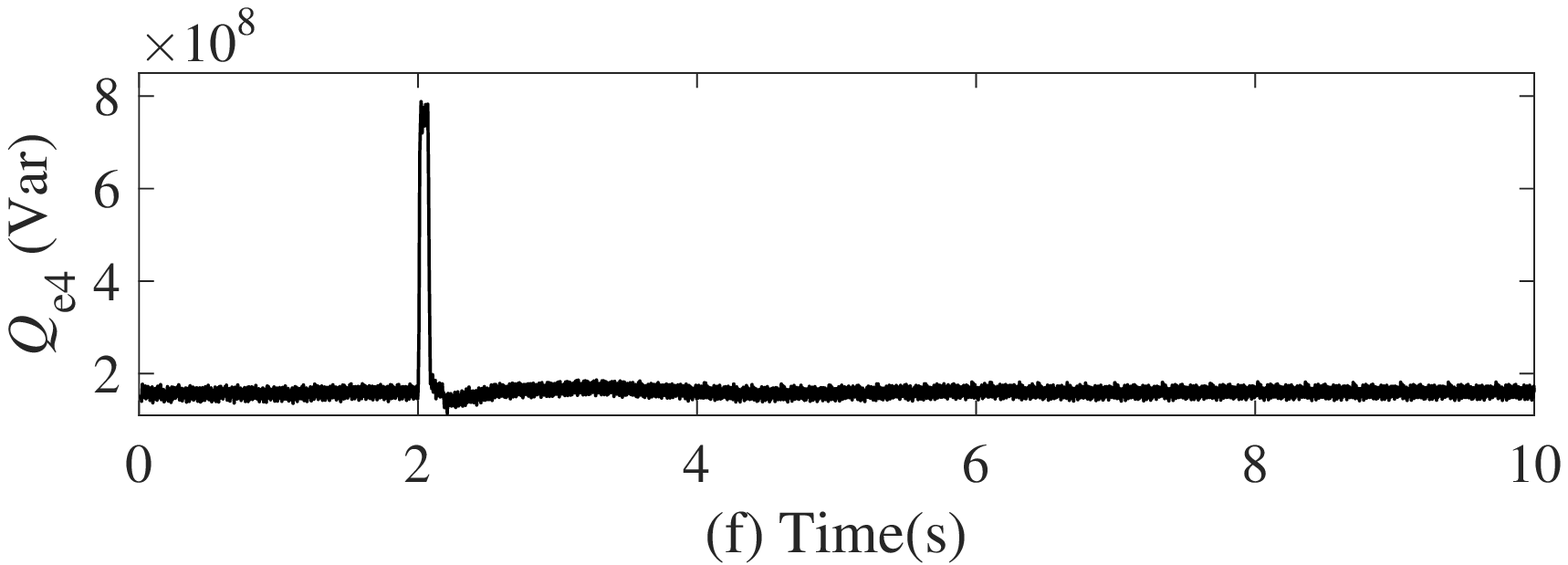}\\
\includegraphics[width=0.485\textwidth]{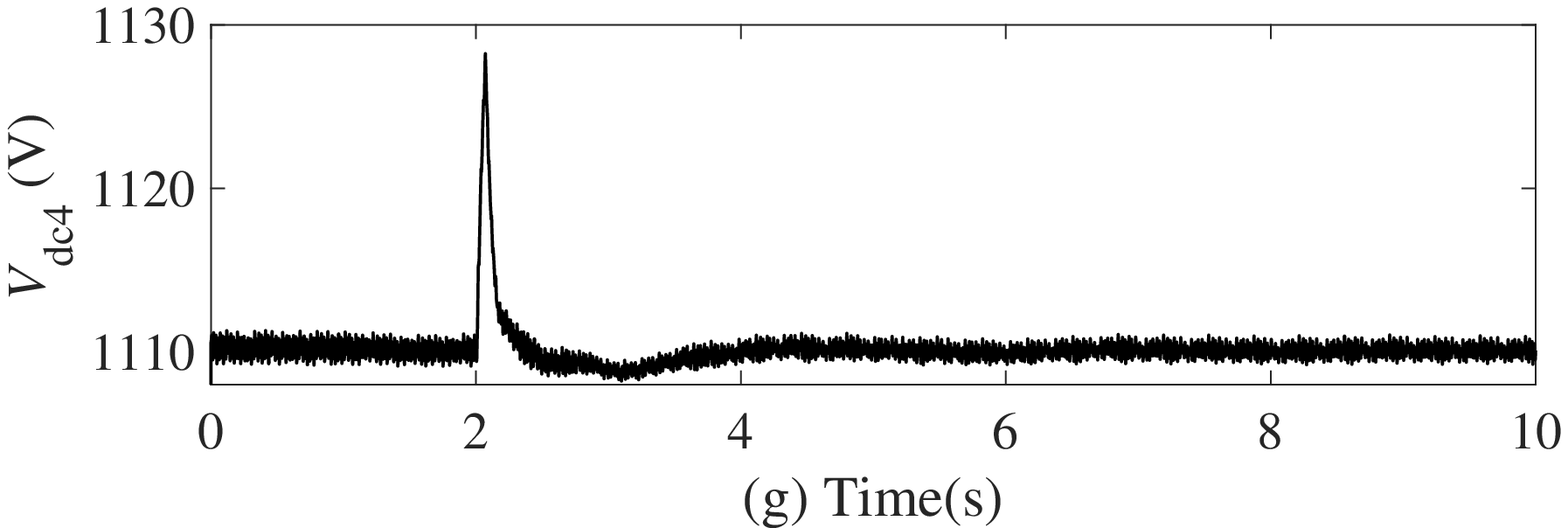}
\caption{Dynamics of WPG$_{4}$ obtained in the case where a three-phase-to-ground fault occurred on the transmission line between bus 7 and bus 8 at $t=2$ s. ((a) Three-phase voltages measured on load bus 7 (b) Three-phase voltages measured on generator bus 4 (c) Active power output of WPG$_{4}$ (d) Power output of the energy storage of WPG$_{4}$ (e) Active power output of the SG of WPG$_{4}$ (f) Reactive power output of WPG$_{4}$ (g) Capacitor voltage of WPG$_{4}$)}
\label{fig_fault1}
\vspace{-0.2cm}
\end{figure}

WPG$_{3}$ and WPG$_{4}$ located on the same side of the FWPS, and they had presented similar dynamics. The active power and reactive power outputs of WPG$_{3}$ also showed step-wise increases as illustrated in Fig. \ref{fig_load_change3} (a) and \ref{fig_load_change3} (b), respectively. Analogously, similar dynamics was observed on WPG$_{2}$ in comparison to that of WPG$_{1}$. The active and reactive power outputs of WPG$_{1}$ are illustrated in Fig. \ref{fig_load_change3} (d) and \ref{fig_load_change3} (e), respectively. The capacitor voltages of WPG$_{3}$ and WPG$_{2}$ synchronized with those of WPG$_{4}$ and WPG$_{1}$, which are illustrated by Fig. \ref{fig_load_change3} (c) and \ref{fig_load_change3} (f), respectively.
\subsection{A Three-phase-to-ground Fault Occurred on the FWPS}
A three-phase-to-ground fault occurred on the transmission line between bus 7 and bus 8 of the FWPS at $t=2$ s. After 4 operating cycles, i.e. 66.7 ms, the fault transmission line was cut off by relay protection devices. Then the transmission line was switched on operation again at $t=2.1667$ s.

Owning to the fault, the voltage of the transmission network of the FWPS dropped. The three-phase voltages measured on load bus 7 dropped to zero during the process when fault happened, which is shown in Fig. \ref{fig_fault1} (a). The voltage drop of the FWPS limited the active power transfer capability of the transmission network. As shown in Fig. \ref{fig_fault1} (c), the active power output of WPG$_{4}$ dropped to less than 400 MW. The capacitor voltage of WPG$_{4}$ then jumped to nearly 1130 V, as depicted in Fig. \ref{fig_fault1} (g). With the effort of the governor, the energy storage of WPG$_{4}$ output less power to the capacitor during the fault process as illustrated in Fig. \ref{fig_fault1} (d). The active power output of the SG of WPG$_{4}$ reduced as well, which is shown in Fig. \ref{fig_fault1} (e). With the exciter of the CVSC system, reactive power output of WPG$_{4}$ was increased to support its PCCB voltage. Consequently, three-phase voltages presented in Fig. \ref{fig_fault1} (b) were obtained on generator bus 4.

\begin{figure}[!t]
\centering
\includegraphics[width=0.48\textwidth]{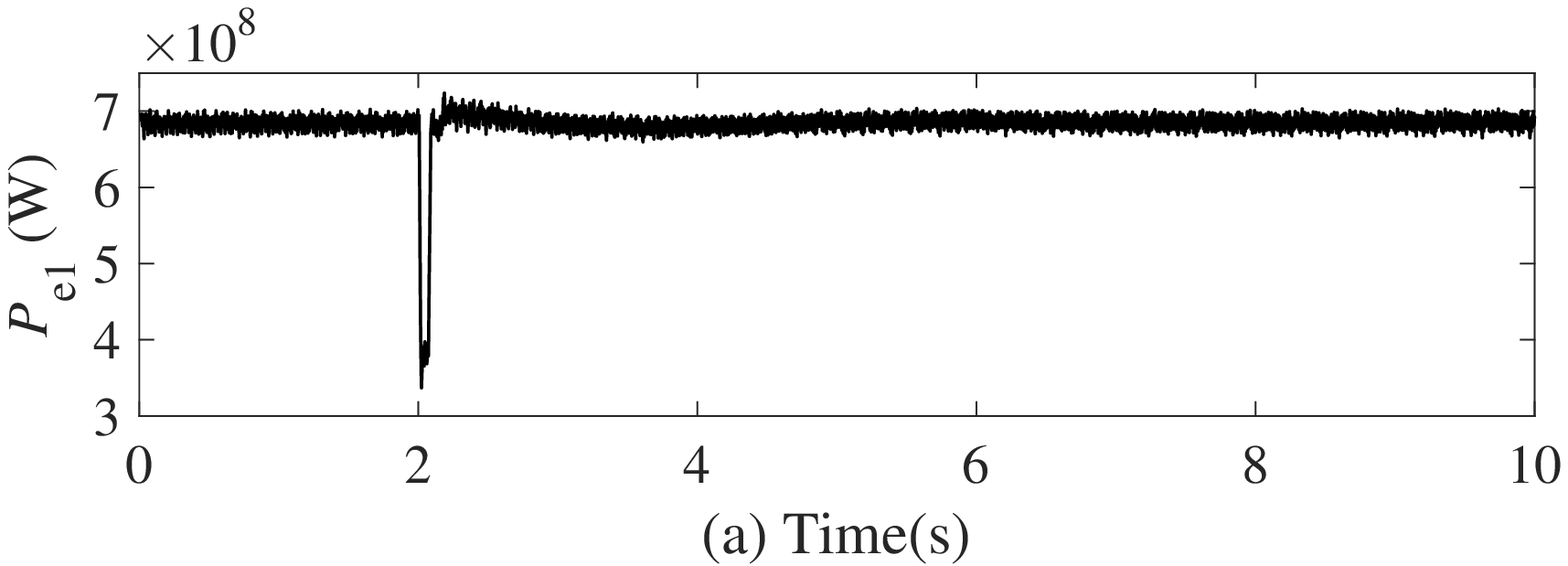}\\
\includegraphics[width=0.48\textwidth]{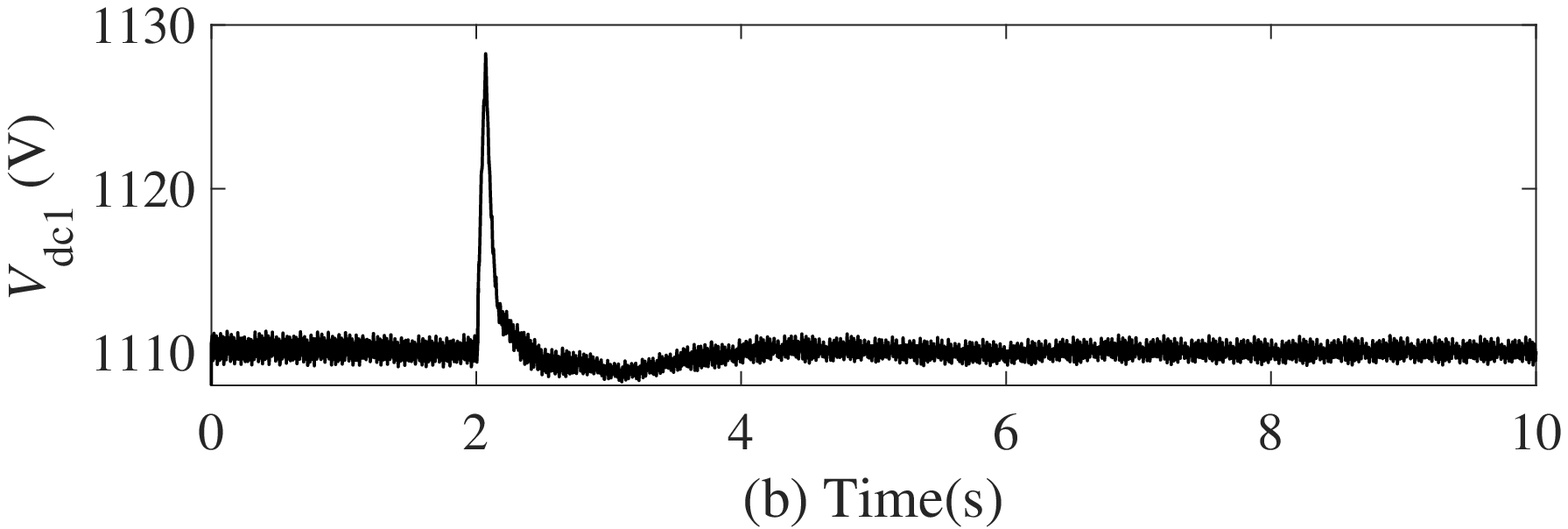}\\
\includegraphics[width=0.48\textwidth]{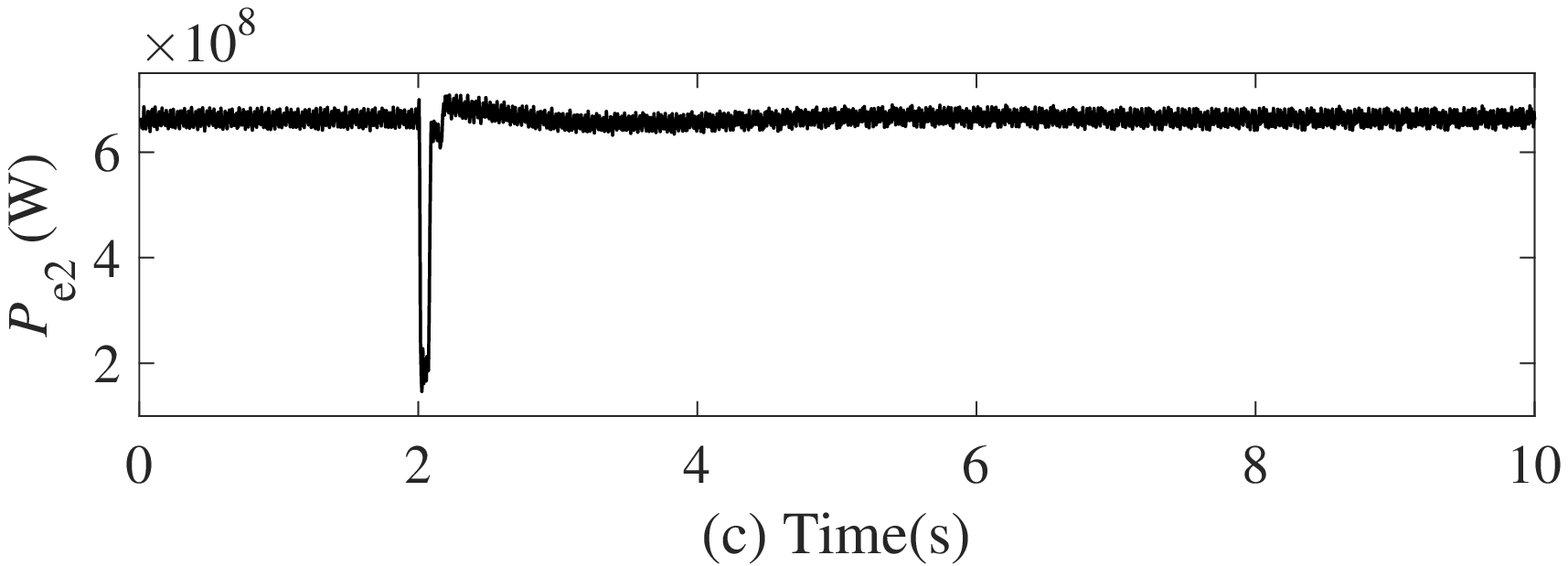}\\
\includegraphics[width=0.48\textwidth]{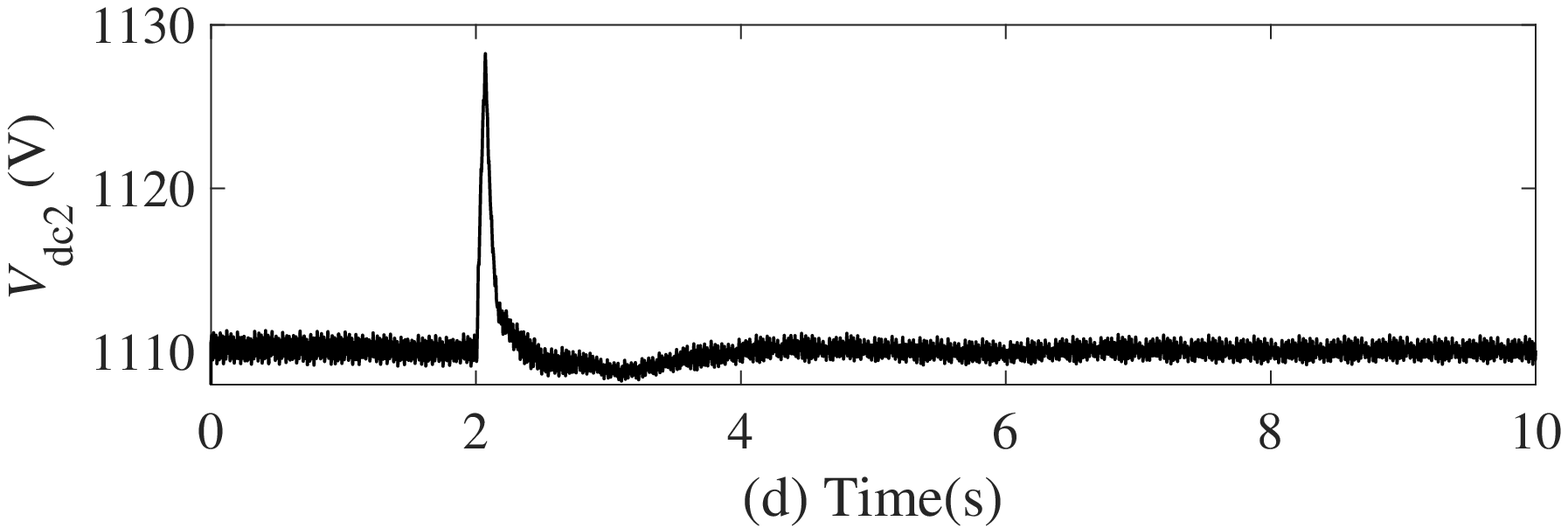}\\
\includegraphics[width=0.48\textwidth]{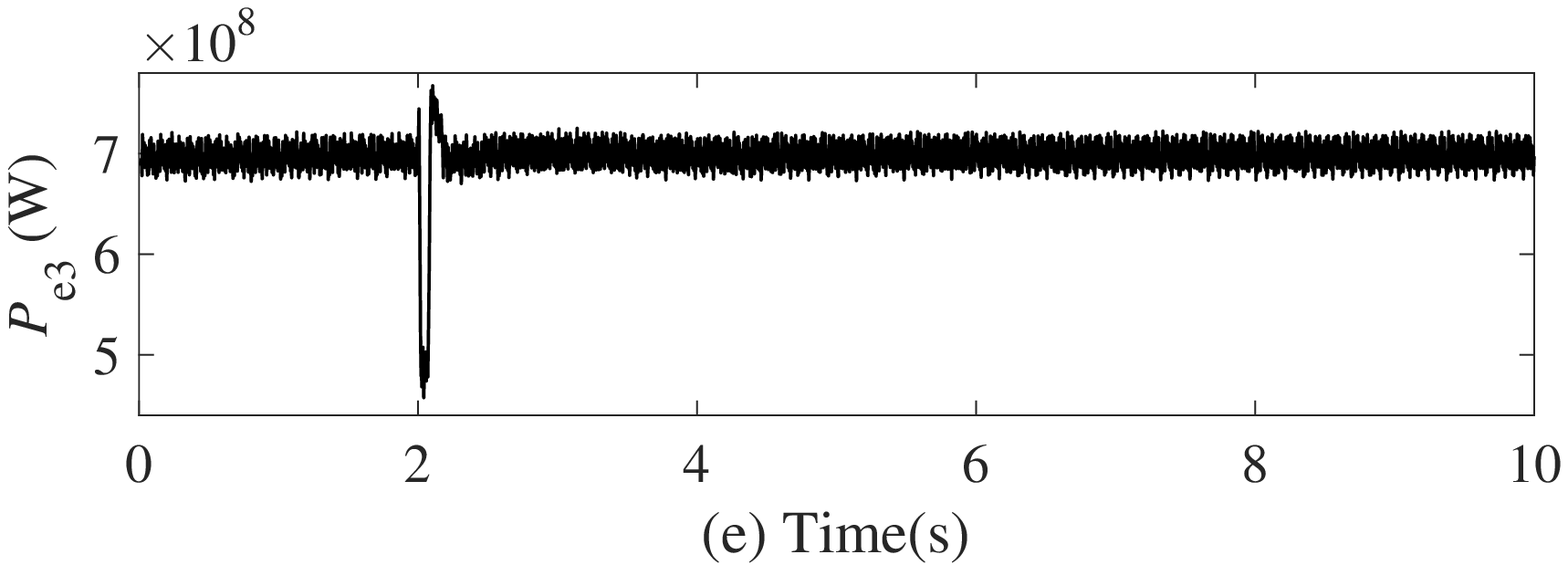}\\
\includegraphics[width=0.48\textwidth]{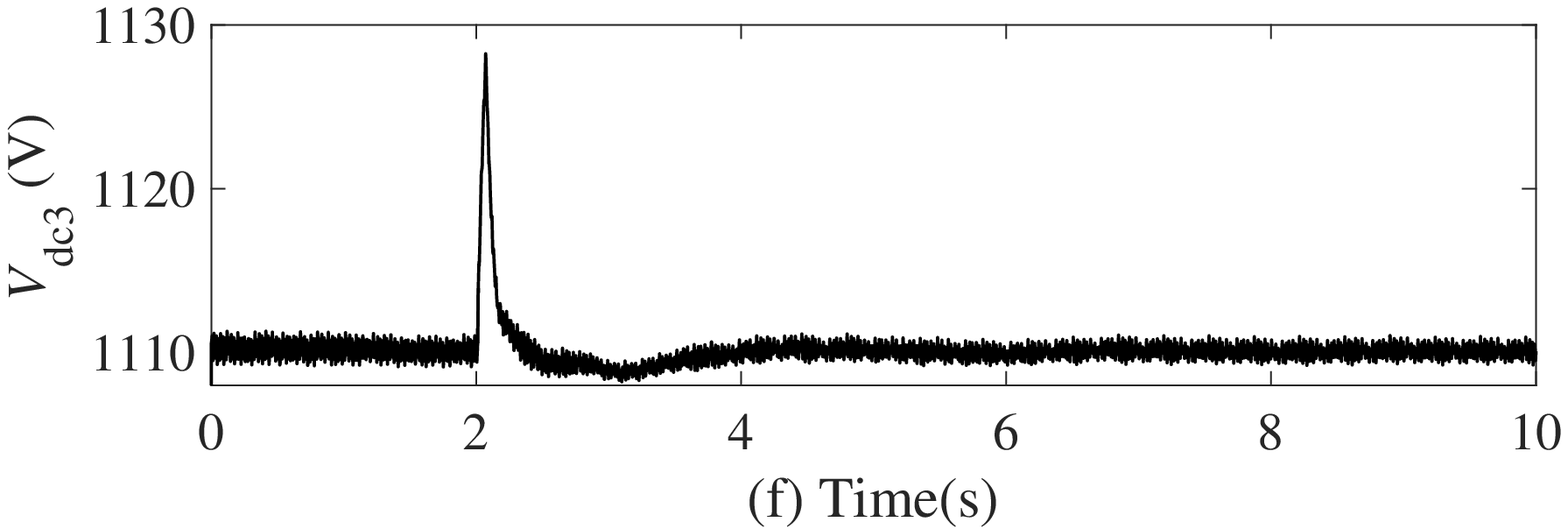}
\caption{Dynamics of WPG$_{1}$, WPG$_{2}$, and WPG$_{3}$ obtained in the case where a three-phase-to-ground fault occurred on the transmission line between bus 7 and bus 8 at $t=2$ s. ((a) Active power output of WPG$_{1}$ (b) Capacitor voltage of WPG$_{1}$ (c) Active power output of WPG$_{2}$ (d) Capacitor voltage of WPG$_{2}$ (e) Active power output of WPG$_{3}$ (f) Capacitor voltage of WPG$_{3}$)}
\label{fig_fault2}
\vspace{-0.6cm}
\end{figure}

Other three WPGs presented similar dynamics with WPG$_{4}$. Fig. \ref{fig_fault2} (a), \ref{fig_fault2} (c), and \ref{fig_fault2} (e) present the dynamics of the active power outputs of WPG$_{1}$, WPG$_{2}$, and WPG$_{3}$, respectively. Active power outputs of all WPGs dropped when the fault happened. Capacitor voltages of WPGs varied in a synchronous manner, which are depicted in Fig. \ref{fig_fault2} (b), \ref{fig_fault2} (d), \ref{fig_fault2} (f), respectively.
\section{Conclusions}\label{sec_conclusion}
This paper has proposed a CVSC system for the regulation of a FWPS. The inverters of WPGs are controlled with the equations of motion of capacitor voltages. With the CVSC system, the FWPS is enabled to operate in the same manner with conventional SG-based power systems.

According to the results of small-signal analysis, the FWPS controlled by the CVSC system is stable in the small. Low-frequency oscillatory modes are found in the rotational speed of wind turbines, and the design of oscillation damping controllers will be presented in our future work.

Simulation results, obtained in the case where a load increase occurred on the FWPS, have verified the capacitor voltage synchronizing performance of WPGs. Capacitors and energy storages of WPGs were able to provide inertial response to the load increase of the external power grid in a coordinated manner, which further enabled the SG of WPGs to offer inertial support to the drop of capacitor voltages. The governor with an active power-capacitor voltage droop characteristic ensured proper active power load sharing between WPGs in the FWPS.

Referring to the simulation results obtained in the case where a three-phase-to-ground fault occurred on the FWPS, the FWPS with the CVSC system is stable in the large. In contrast to the original four-generator two-area system, few low-frequency oscillations were observed in the dynamics of active power outputs and capacitor voltages of WPGs. The FWPS presented different transient characteristics in terms of high-frequency harmonic oscillations in both the tested cases, which is the inspiration for future works considering the harmonic stability of the FWPS.

\bibliographystyle{IEEEtran}
\bibliography{Ref}
\appendices
\section{Nomenclature}\label{appen_nomenclature}
\footnotesize
\begin{description}
  \item[$\omega_{\mathrm{r}i}$] Rotor speed of the SG of WPG$_{i}$
  \item[$\omega_{\mathrm{r}\_\mathrm{ref}i}$] \quad Reference of $\omega_{\mathrm{r}i}$
  \item[$K_{\mathrm{p}\_\mathrm{pitch}}$] \quad\; Proportional gain of the rotor speed control loop of wind turbine governors
  \item[$\beta_i$] Pitch angle of the turbine blades of WPG$_{i}$
  \item[$P_{\mathrm{me}i}$] Power input to the capacitor of WPG$_{i}$
  \item[$P_{\mathrm{me}\_\mathrm{ref}i}$] \quad\; Reference of $P_{\mathrm{me}i}$
  \item[$P_{\mathrm{in}i}$] Active power output of the SG of WPG$_{i}$
  \item[$P_{\mathrm{in}\_\mathrm{ref}i}$]\quad The reference of $P_{\mathrm{in}i}$
  \item[$K_{\mathrm{p}\_\mathrm{comp}}$] \quad\; Proportional gain of the proportional-integral (PI) controller of the power control loop of wind turbine governors
  \item[$K_{\mathrm{i}\_\mathrm{comp}}$] \quad\; Integral gain of the PI controller of the power control loop of wind turbine governors
  \item[$|\Psi_{i}|$] Magnitude of the stator flux of the SG of WPG$_{i}$
  \item[$|\Psi_{i}|_{\mathrm{ref}}$] \quad Reference of $|\Psi_{i}|$
  \item[$E_{\mathrm{f}i}$] Excitation voltage of the SG of WPG$_{i}$
  \item[$K_{\mathrm{p}\_\mathrm{field}}$] \quad Proportional gain of the PI controller of the exciter of SGs
  \item[$K_{\mathrm{i}\_\mathrm{field}}$] \quad Integral gain of the PI controller of the exciter of SGs
  \item[$C$] Capacity of the capacitor of WPGs
  \item[$V_{\mathrm{dc}i}$] Capacitor voltage of WPG$_{i}$
  \item[$V_{\mathrm{dc}\_\mathrm{nom}}$]\quad Nominal capacitor voltage of WPGs
  \item[$K_{\mathrm{pg}1}$] Proportional gain of the governor of WPGs
  \item[$K_{\mathrm{pg}2}$] Proportional gain of the governor of WPGs
  \item[$K_{\mathrm{pg}3}$] Proportional gain of the governor of WPGs
  \item[$I_{\mathrm{s}i}$] Current injected by the energy storage of WPG$_{i}$
  \item[$I_{\mathrm{s}\_\mathrm{ref}i}$]\quad Reference of $I_{\mathrm{s}i}$
  \item[$L_{\mathrm{boost}}$] \quad Inductance of the boost converter connected outside the rectifier
  \item[$D_{\mathrm{duty}}$] \quad Duty cycle of the boost converter of WPGs
  \item[$\Delta\theta_{i}$] Output of the phase angle control loop of the CVSC system of WPG$_{i}$
  \item[$v'_{\mathrm{d}i}$] d-axis reference voltage of the inverter of WPG$_{i}$
  \item[$v'_{\mathrm{q}i}$] q-axis reference voltage of the inverter of WPG$_{i}$
  \item[$\omega_{i}$] Rotational speed of the inverter voltage vector of WPG$_{i}$
  \item[$m_{i}$] Modulation index of the inverter of WPG$_{i}$
  \item[$f_{\mathrm{n}}$] Nominal frequency of the FWPS
  \item[$K_{\mathrm{a}}$] Gain of the phase angle control loop of the CVSC system
  \item[$K_{\mathrm{e}}$] Gain of the exciter of the CVSC system
  \item[$V_{\mathrm{t}i}$] Magnitude of the three-phase voltages measured on the PCCB of WPG$_{i}$
  \item[$V_{\mathrm{t}\_\mathrm{ref}i}$] Reference of $V_{\mathrm{t}i}$
  \item[$L_{\mathrm{filter}}$] Inductance of the $RL$ filter connected outside the inverter of WPGs
  \item[$R_{\mathrm{filter}}$] Resistance of the $RL$ filter connected outside the inverter of WPGs
  \item[$P_{\mathrm{mn}i}$] Nominal mechanical power of WPG$_{i}$
  \item[$P_{\mathrm{n}i}$] Nominal electrical power of WPG$_{i}$
  \item[$V_{\mathrm{n}}$] Nominal line-to-line voltage of the stator of the SG of WPGs
  \item[$V_{\mathrm{tn}}$] Nominal line-to-line voltage of the PCCB of WPGs
  \item[$x_{\mathrm{d}}$] d-axis synchronous reactance of the SG of WPGs
  \item[$x'_{\mathrm{d}}$] d-axis transient reactance of the SG of WPGs
  \item[$x''_{\mathrm{d}}$] d-axis sub-transient reactance of the SG of WPGs
  \item[$x_{\mathrm{q}}$] q-axis synchronous reactance of the SG of WPGs
  \item[$x''_{\mathrm{q}}$] q-axis sub-transient reactance of the SG of WPGs
  \item[$x_{\mathrm{l}}$] Stator leakage reactance of the SG of WPGs
  \item[$T'_{\mathrm{d0}}$] d-axis transient open-circuit time constant of the SG of WPGs
  \item[$T''_{\mathrm{d0}}$] d-axis sub-transient open-circuit time constant of the SG of WPGs
  \item[$T''_{\mathrm{q0}}$] q-axis sub-transient open-circuit time constant of the SG of WPGs
  \item[$R_{\mathrm{s}}$] Stator resistance of the SG of WPGs
  \item[$p$] Number of pole pairs of the SG of WPGs
  \item[$\Psi_{\mathrm{d}i}$] d-axis stator flux of the SG of WPG$_{i}$
  \item[$\Psi'_{\mathrm{kq}i}$] q-axis transient flux of the damper windings of the SG of WPG$_{i}$
  \item[$\Psi'_{\mathrm{kd}i}$] d-axis transient flux of the damper windings of the SG of WPG$_{i}$
  \item[$\omega_{\mathrm{t}i}$] Mechanical rotational speed of the wind turbine of WPG$_{i}$
  \item[$\Psi'_{\mathrm{f}i}$] Transient flux of the field windings of the SG of WPG$_{i}$
\end{description}

\section{Parameters of the Full-Scale WPGs of the FWPS}\label{appen_parameter}
\begin{table}[htbp]
\centering
\caption{Parameters of full-scale WPGs}
\begin{tabular}{p{1cm}p{1.2cm}p{1cm}p{1.2cm}p{1cm}p{1.2cm}}
\hline
Parameter & Value & Parameter & Value & Parameter & Value \\
\hline
$P_{\mathrm{mn}i}$ & 800 MW & $P_{\mathrm{n}}$ & 889 MW & $V_{\mathrm{n}}$ & 730 V \\
$f_{\mathrm{n}}$ & 60 Hz & $V_{\mathrm{tn}}$ & 575 V & $L_{\mathrm{filter}}$ & 0.15 p.u. \\
$R_{\mathrm{filter}}$ & 0.003 p.u. & $V_{\mathrm{dc}\_\mathrm{nom}}$ & 1110 V & $C$ & 36 F \\
$L_{\mathrm{boost}}$ & 0.0012 H & $K_{\mathrm{p}\_\mathrm{pitch}}$ & 15 & $K_{\mathrm{p}\_\mathrm{comp}}$ & 1.5 \\
$K_{\mathrm{i}\_\mathrm{comp}}$ & 6 & $K_{\mathrm{p}\_\mathrm{field}}$ & 10 & $K_{\mathrm{i}\_\mathrm{field}}$ & 20 \\
$K_{\mathrm{pg1}}$ & 30 & $K_{\mathrm{pg2}}$ & 15 & $K_{\mathrm{pg3}}$ & 0.1 \\
$p$ & 1 & $K_{\mathrm{a}}$ & 10 & $K_{\mathrm{e}}$ & 0.2 \\
$D_{\mathrm{duty}}$ & 0.19 & $x_{\mathrm{d}}$ & 1.305 p.u. & $x'_{\mathrm{d}}$ & 0.296 p.u. \\
$x''_{\mathrm{d}}$ & 0.252 p.u. & $x_{\mathrm{q}}$ & 0.474 p.u. & $x''_{\mathrm{q}}$ & 0.243 p.u. \\
$x_{\mathrm{l}}$ & 0.18 p.u. & $T'_{\mathrm{d0}}$ & 4.49 s & $T''_{\mathrm{d0}}$ & 0.0681 s \\
$T''_{\mathrm{q0}}$ & 0.0513 & $R_{\mathrm{s}}$ & 0.006 p.u. &  & \\
\hline
\end{tabular}
\end{table}
\begin{IEEEbiography}[{\includegraphics[width=1in,height=1.25in,clip,keepaspectratio]{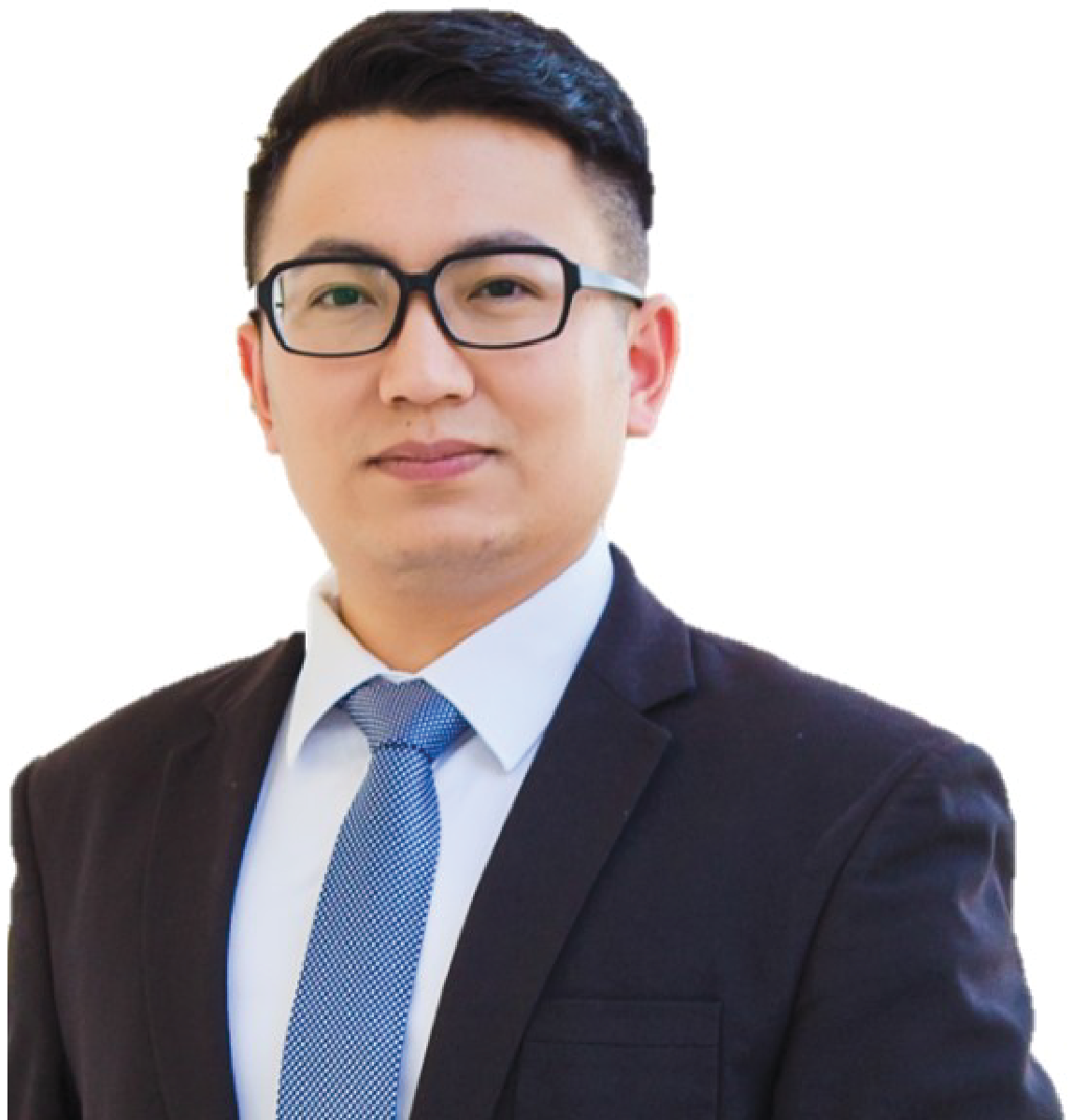}}]{Yang Liu}
\footnotesize{(CSEE: M'18, IEEE: M'18) received a B.E. degree and a Ph.D degree in Electrical Engineering from South China University of Technology (SCUT), Guangzhou, China, in 2012 and 2017, respectively. He is currently an Associate Professor in the School of Electric Power Engineering, SCUT. His research interests include the areas of power system stability analysis and control, control of wind power generation systems, and nonlinear control theory. He has authored or co-authored more than 30 peer-reviewed SCI journal papers, and a monograph published by Springer Nature named as ``Switching Control of Large-Scale Complex Power Systems-Theory and Applications".}
\end{IEEEbiography}
\end{document}